\documentclass[aps,pra,amsmath,notitlepage,nofootinbib,twocolumn,superscriptaddress,10pt]{revtex4-2}
\pdfoutput=1
\usepackage[utf8]{inputenc}
\usepackage[english]{babel}
\usepackage{microtype}
\usepackage[normalem]{ulem}

\usepackage{graphicx}
\usepackage{dcolumn}
\usepackage{bm}
\usepackage{amsmath,amssymb,amsfonts}
\usepackage{booktabs}

\usepackage{braket}
\usepackage{dsfont}
\usepackage{xcolor}
\usepackage{dsfont}
\usepackage{mathtools}

\usepackage{tikz}

\usepackage[breaklinks=true,colorlinks=true,linkcolor=teal,urlcolor=teal,citecolor=teal]{hyperref}
\usepackage{orcidlink}

\let\originalleft\left
\let\originalright\right
\renewcommand{\left}{\mathopen{}\mathclose\bgroup\originalleft}
\renewcommand{\right}{\aftergroup\egroup\originalright}

\begin{document}

\title{Multiphoton Hong–Ou–Mandel Interference Enables Superresolution of Bright Thermal Sources}

\author{Aiman Khan\,\orcidlink{0009-0006-0697-8343}}
\thanks{aimankhan1509@protonmail.com}
\affiliation{Manufacturing Metrology Team, Faculty of Engineering, University of Nottingham, Nottingham NG7 2RD, United Kingdom}
\affiliation{School of Mathematics and Physics, University of Portsmouth, Portsmouth PO1 3QL, UK.}
\affiliation{Quantum Science and Technology Hub, University of Portsmouth, Portsmouth P01 3QL, UK.}

\author{Danilo Triggiani\,\orcidlink{0000-0001-8851-6391}}
\affiliation{Dipartimento Interateneo di Fisica, Politecnico di Bari, 70126, Bari, Italy}
\affiliation{Istituto Nazionale di Fisica Nucleare (INFN), Sezione di Bari, 70126 Bari, Italy}

\author{Vincenzo Tamma\,\orcidlink{0000-0002-1963-3057}}
\thanks{vincenzo.tamma@port.ac.uk}
\affiliation{School of Mathematics and Physics, University of Portsmouth, Portsmouth PO1 3QL, UK.}
\affiliation{Quantum Science and Technology Hub, University of Portsmouth, Portsmouth PO1 3QL, UK.}
\affiliation{Institute of Cosmology and Gravitation, University of Portsmouth, Portsmouth PO1 3FX, UK.}

\date{\today}

\begin{abstract}
We present a quantum optical scheme for imaging transversely displaced thermal sources of arbitrary intensities by employing multiphoton interference with a reference single-photon Fock state at a beamsplitter. Obtaining an analytical form for transverse momenta-resolved $L$-photon probabilities in either output, we show via Fisher information analysis that separation estimators built using interference sampling of multiphoton events exhibit significantly enhanced precision vis-\`{a}-vis existing imaging schemes over a wide range of separations and brightness. Even-photon-number coincidences exhibit constant precision in the sub-Rayleigh regime, demonstrating quantum superresolution of our scheme beyond the diffraction limit. For sources emitting on average $N_s\sim1$ photon per frame~(such as in IR emission of thermal sources), precision bounds for our scheme scale linearly in $N_s$, exemplifying an enhanced precision of estimators in relation to weak sources $N_s\ll1$, and matching the ultimate quantum scaling. Finally, transverse momenta resolution in the Fourier plane produces finite imaging precisions for intermediate and large source separations using coarse pixel sizes of order $\delta y\sim100\,\mu \mathrm{m}$ for exemplary image spot sizes $\sigma_x \sim 0.1\, \mu \mathrm{m}$, in contrast with existing schemes of diffraction-limited direct imaging and superresolved inversion interferometric imaging that are severely degraded by coarse pixel sizes and have limited use. Combining the relatively straightforward sensing operation of Hong-Ou-Mandel interferometers with multiphoton coincidence detection of arbitrarily bright thermal sources and inner variable resolution of transverse photonic momenta, our scheme offers a robust alternative to non-invasive single-particle tracking and imaging of bright sources in nanoscopic chemical and biological systems. 
\end{abstract}

\maketitle

\section{Introduction}

Since its inception, remarkable progress has been made in employing Hong-Ou-Mandel~(HOM) interference to sense photonic properties~\cite{hong1987measurement,lyons2018attosecond,bouchard2020two}. Exploiting the multiphoton quantum interference of identical paths that two completely indistinguishable single-photon inputs to a balanced beamsplitter~(BS) may take, the HOM effect predicts that the two photons ``bunch" at the output, while never being simultaneously detected in different output arms~(``anti-bunching"). The precision sensing is based on the fact that any distinguishability in the input photons~(induced by differential properties such as time delays, path lengths, or polarisations) would curtail this interference, making possible again for photons to be detected simultaneously in the two arms. This produces then, as function of the distinguishability, the eponymous ``HOM dip" in two-photon coincidence, offering large signal-to-noise ratios against a reduced background signal. HOM interferometers have been used as time delay sensors with attosecond resolutions~\cite{lyons2018attosecond}, for fluorophore lifetime sensing~\cite{lyons2023fluorescence}, and also in interferometric nonlinear spectroscopy setups that sense matter susceptibilities~\cite{dorfman2021hong}, amongst myriad other photonic applications.

Originally proposed for two-photon interference, the HOM effect was recently extended to multiphoton interference at BS between arbitrary optical states and otherwise identical non-classical states of odd parity\cite{alsing2022extending,alsing2025examination}, where it was shown that the output photon number distribution in the two arms contains no counts for an equal number of photons detected in each arm, a striking generalisation of the HOM dip for multiphoton events. This then raises the question of whether there is an analogous sensing advantage for bunched light states~(where multiphoton events are expected to be important) by utilising this extended HOM~(eHOM) effect, characterized by a valley in photon number distribution rather than a coincidence dip~\cite{birrittella2012multiphoton,alsing2022extending}. 

In this respect, one prominent sensing task confronting quantum optics is the incoherent imaging of optical sources beyond the diffraction limit~\cite{tsang2019resolving,moreau2019imaging}. Couching the task of resolving two equally intense faint sources as a quantum estimation problem, Tsang and Nair~\cite{tsang2016quantum} demonstrated that the resolution limits of Abbe and Rayleigh for `direct imaging' that measures intensity as a function of the transverse position in the collection plane can be superseded by several orders in alternative, quantum-optimised measurement bases. This epochal result has since precipitated a wide search for robust optical techniques that can realise this theoretical quantum superresolution in incoherent imaging, chief amongst them being spatial demultiplexing~(SPADE)~\cite{tsang2017subdiffraction,rouviere2024ultra,len2020resolution}, superresolved position localisation by inversion of coherence along an edge~(SPLICE)\cite{tham2017beating}, and superlocalisation by image inversion~(SLIVER) interferometry~\cite{nair2016interferometric,nair2016far}. Remarkably, these `quantum' superresolution techniques differ from the more conventional superresolution imaging strategies~(such as photoactivated localisation microscopy~(PALM) and stimulated emission depletion~(STED) microscopy~\cite{betzig2015nobel,hell2015nobel}) in that they do not require \emph{any} modification or control over the emission properties of the source, rather relying on the quantum nature of the emitted light. This feature holds enormous promise for diverse applications in astronomical interferometry~\cite{sajjad2024quantum,padilla2024quantum}, remote sensing~\cite{kose2022quantum}, as well as non-invasive bioimaging and sensing~\cite{li2013quantum}, where patterned blinking illumination of the sources is either inaccessible or undesirable.

Although the original demonstration of quantum superresolution in \cite{tsang2016quantum} made the assumption of faint sources that emit at most one photon in a single frame~(which is a reasonable assumption for stellar photons), it is possible to extend this result to brighter sources.  For the paradigmatic problem of resolving two incoherent thermal sources, except now of arbitrary strength $N_s$, it has been shown~\cite{lupo2016ultimate,tsang2019resolving} that the ultimate estimation precision scales as $O(N\times N_s)$~(where $N$ is the number of frames), and that quantum superresolution beyond the diffraction limit persists just as for weak thermal light~\cite{tsang2016quantum}. Consequentially, brighter optical sources may be resolved in fewer experimental runs, as the standard quantum limit~(SQL) scaling for independent and identically distributed sources is now amplified by the intensity $N_s$. Bright sources with bunched, super-Poissonian statistics are routinely encountered in astronomical imaging at longer wavelengths~(infrared and microwave regions)~\cite{schlawin2023jwst,sallum2024jwst}, in laboratory demonstrations of pseudothermal light using laser sources~\cite{estes1971scattering}, as well as for scattered coherent light~\cite{mandel1995optical}.

At the same time, there are practical challenges to imaging bright thermal sources in quantum imaging setups. In SPADE, the foremost of this newly developed class of imaging techniques, resolution limits are deeply impacted by large number of frames $N$ when sources are separated by sub-Rayleigh distances -- in fact, the scaling of the estimator precision has been forecasted to scale as $O(\sqrt{N})$, worse than shot noise, for $N>>1$~\cite{gessner2020superresolution}. While this determination was made for weak thermal light sources, one can reasonably expect this effect to persist and worsen for bunched photons detected in orthogonal modes. The other near-optimal candidate for bright source quantum imaging, SLIVER~\cite{nair2016far,nair2016interferometric}, suffers from path difference instability in the two arms of the Mach-Zehnder interferometer~(MZI) that is central to the imaging setup. This path difference, which must be smaller than the wavelength of the emitted light, is unsurprisingly hard to control in practical microscopies~\cite{wicker2009characterisation,tang2016fault} to the required precision. In addition, unavoidable aberrations as well as imperfect splitting between the two arms of the MZI employed to separate the primary modes from the derivative modes, also deeply affect sub-Rayleigh resolutions making experimental implementations problematic~\cite{schodt2023tolerance}. Ingenious methods have been identified in experiments to overcome some of these challenges, such as common-path interferometer to eliminate the need for alignment~\cite{larson2019common}, polarisation filtering that accounts for the dipole nature of fluorescent point sources for which wide numerical apertures must be applied~\cite{mitchell2024quantum}, as well as and feedback-enabled phase locking to stabilise optical path difference ~\cite{aiello2025sub}.  However, implementations beyond proof-of-concept have remained elusive 

On the whole, what these experimental difficulties illustrate is the fragile nature of the quantum advantage in the sub-Rayleigh regime, where experimental imperfections and noise have an outsize impact~\cite{lupo2020subwavelength,oh2021quantum}. This demonstrates the critical need for \emph{simpler} imaging setups that can achieve superresolution without involving multi-stage manipulation of the image state. 

In this article, we demonstrate that this can be achieved in a scheme that leverages multiphoton  interference of arbitrary strength thermal sources with a reference single-photon state at the BS in order to image the sources beyond the Rayleigh limit. Rooted in the relatively simple and extensively employed HOM setup, our technique directly exploits the multiphoton correlations that characterise intense, super-Poissonian light which are increasingly accessible in experiments with pixelated camera arrays of single-photon detectors, representing a realistic and easily adaptable quantum optical scheme to image thermal electromagnetic sources of arbitrary strengths. Eliminating both the need for path alignment as for SLIVER, and possibility of crosstalk that affects resolution limits in SPADE, we show that significant quantum enhancement may be achieved by harnessing the power of multiphoton interference at the BS. Advances in large-format single-photon avalance photodiode~(SPAD) and superconducting nanowire single-photon detectors~(SNSPDs) imaging arrays have enabled spatially resolved, time-tagged multiphoton coincidence detection with high efficiency and picosecond timing resolution, and have already demonstrated number-resolved photon counting across many pixels, placing the detection requirements of our scheme well within current experimental capabilities~\cite{incoronato2021multi,zhao2017single,cheng2023100}.

Our main results are summarised as follows. First, we find that separation estimators for our imaging scheme exceed diffraction limits for \emph{all} even-order photon coincidences, directly surpassing the HOM sensing scheme in \cite{muratore2025superresolution} for weak thermal sources that uses only two-photon coincidences. Secondly, by further resolving transverse momenta of the output photons in the far-field, we demonstrate the emergence of multiphoton beats beyond Rayleigh separations which imparts our imaging scheme near constant precision over both sub- and super-Rayleigh source separations. This resolution of transverse momenta in the Fourier plane allows for the use of inexpensive coarse pixel sizes as much one order of magnitude larger than the spot size without affecting precision, offering advantage against direct imaging and SLIVER where such coarse-grained pixelated imaging offer practically zero precision. Lastly, we also demonstrate that for source brightnesses $N_s \sim 1$ appropriate for imaging in the IR spectrum, the scheme precision scales as $O(N_s)$, matching the ultimate quantum precision found in \cite{lupo2016ultimate,nair2016far}.

The article is structured as follows :  in Section \ref{sec:setup}, we introduce the quantum-optical formalism of the imaging setup; in Section \ref{sec:imaging}, we  obtain an analytical form for the $L$-photon coincidence probabilities at the heart of the imaging scheme; in Section \ref{sec:metrology}, we evaluate and study precision bounds set by Fisher informations (FIs) for multiphoton orders, where we show that superresolution may be obtained in all even orders for our scheme, and that the transverse momenta resolution allows superior imaging precision relative to other quantum imaging schemes for varying separation and brightness; finally in Section \ref{sec:conclusions}, we conclude with a summary of planned future work.

\section{Quantum Optical Model}\label{sec:setup}
Consider two point quasimonochromatic thermal sources, of arbitrary~(but equal) intensities characterised by average photon number $N_s$, and spatially displaced by distance $s$ on a plane transverse to photon propagation. For simplicity, we assume that they are localised along a single line~(labelled the $x$-axis) at positions $x_{\pm}=\pm s/2$, implicitly assuming that the centroid of the distribution is known, $\bar{x} = 0$. 

Using only the minimal assumptions of unit magnification and spatial invariance for the imaging device, the quantum state of the electromagnetic field at the image field may be expressed as the coherent state decomposition\,\cite{mandel1995optical}:
\begin{equation}
   \rho_d(s) = \int_{\mathbb{C}} d^2A_+d^2A_-~P_{N_s}(A_+,A_-)\,\ket{\psi_{A,s}}\bra{\psi_{A,s}},
\end{equation}
with normally-distributed coherent source amplitudes $P_{N_s}(A_+,A_-) = \frac{1}{(\pi N_s)^2}\mathrm{exp}\left[\frac{|A_+|^2+|A_-|^2}{2N_s}\right]$ corresponding to the thermal nature of the source fields, and the image field amplitude
\begin{equation}
    \psi_{A,s}(x) = A_+\psi(x-s/2) + A_-\psi(x+s/2)
\end{equation}
fixed by both the source distribution as well as the nature of the imager\,\cite{nair2016far}. The function $\psi(x)\in \mathds{C}$ is identified as the point spread function~(PSF), and characterises the distortion accrued in the propagation of the electromagnetic field through the imager. In typical imaging scenarios, the non-zero overlap of the PSFs corresponding to the two sources, $\psi(x-s/2)$ and $\psi(x+s/2)$, 
\begin{equation}
    \delta(s) = \int dx\, \psi^*(x-s/2) \psi(x+s/2)
\end{equation}
is the primary source of complexity in resolving the two sources, and sets the Rayleigh limit of direct imaging resolution~\cite{born2013principles,zhou2019modern}. In the quantum-optical sense, $\delta>0$ introduces correlations in the direct spatial mode bases corresponding to each source, yielding as a consequence a correlated thermal image state $\rho_d$. This correlated-ness can be remedied using the basis transformation~\cite{tsang2016quantum}, 
\begin{equation}
    \psi_{\pm}(x) = \frac{1}{\sqrt{2(1\pm\delta)}}\,[\psi(x+s/2) \pm \psi(x-s/2)],
\end{equation}
and corresponding transformation of $P$-amplitudes $S = A_+ + A_-; ~~ D = A_+-A_-$, resulting in the following uncorrelated image state composed of tensor product of thermal states but with unequal intensities:
\begin{equation}\label{eq:rhod}
    \rho_d = \rho^{+}_{\mathrm{th}}[M_{+}]\otimes \rho^{-}_{\mathrm{th}}[M_{-}],
\end{equation}
where $\rho^{\pm}_{\mathrm{th}}[M] = \sum_m \frac{1}{m!}\frac{M^m}{(M+1)^{m+1}}  (a_{\pm}^{\dag})^m\ket{\mathrm{vac}}\bra{\mathrm{vac}}(a_{\pm})^m$, with
\begin{equation}
    a_{\pm} = \frac{1}{\sqrt{2(1\pm\delta)}}\int dx_1  \bigg[ \psi(x+s/2) \pm \psi(x-s/2)  \bigg] a_x,
\end{equation}
and $M_{\pm} = N_s(1\pm\delta)$. This can be more succinctly expressed, for our purposes later in the paper, as the following Fock-state decomposition\cite{lupo2016ultimate}:
\begin{equation}\label{eq:lupopirandolaform}
    \rho_{d} =  p_0\sum_{m,n}~p_m^{+} p_n^{-}\ket{m;n}\bra{m;n},
\end{equation}
where $p_0 = 1/[(1+N_s)^2-N_s^2\delta^2]$, and $p_m^{\pm} = \frac{M_{\pm}^m}{(M_{\pm}+1)^{m}}$. Intuitively, the translation from the source field~(where the sources are equally bright and uncorrelated) to the image field is encoded as the overlap-dependent luminosities of the (also uncorrelated) two-mode field state created by mode operators $a_{\pm}$.

\section{Imaging via Multiphoton Interference with Single-Photon Fock State}\label{sec:imaging}

\begin{figure}
    \centering
    \includegraphics[width=0.5\textwidth]{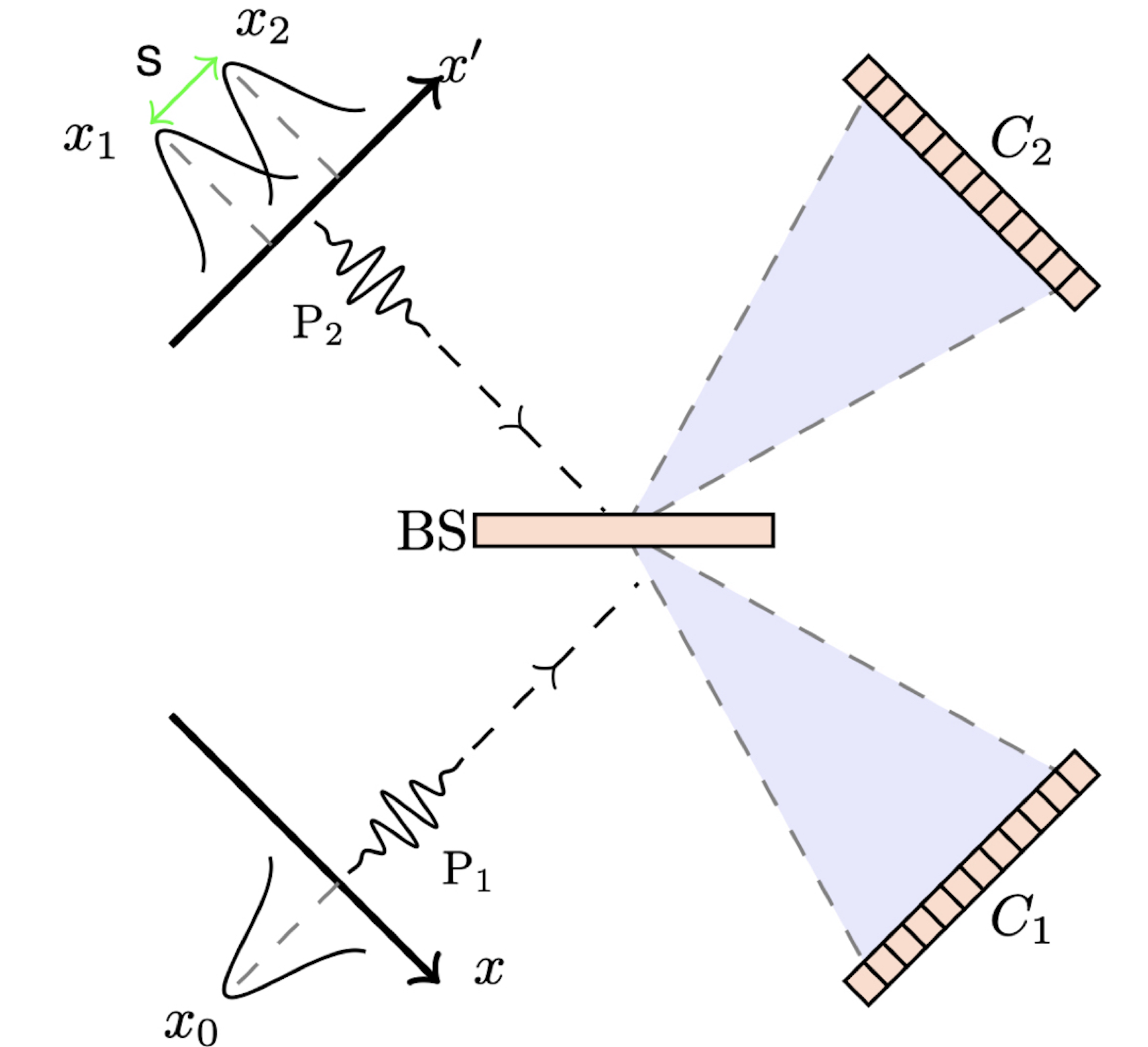}
    \caption{\emph{Imaging scheme setup}: Image state $\rho_d(s)$ corresponding to thermal source distribution interferes at the balanced BS with a single-photon reference state whose wavepacket center coincides with the image centroid. After propagating to the far field, cameras $C_1$ and $C_2$ composed of single-photon detector arrays record in coincidence the spatial positions of the impinging photons, resolving the corresponding transverse momenta.}
    \label{fig:setup}
\end{figure}

We will now describe our imaging scheme that employs multiphoton quantum interference in order to resolve the relative positions of the sources encoded onto the image field state $\rho_d(s)$. In order to do so, we have the image state impinge on one of the two input modes (labelled $S=1)$ of a balanced beamsplitter~(BS), while the other input mode~(labelled $S=0$) is illuminated by a reference single-photon wavepacket centred at spatial transverse position  $x_0$, 
\begin{equation}
    \ket{\psi_0} = \int dx~\psi(x-x_0)a_0(x)\ket{\mathrm{vac}}_0.
\end{equation}
In subsequent discussion, we will always assume $x_0=\bar{x} = 0$ so that the wavepacket centre of the reference state coincides with the centroid of the image distribution, corresponding to the fact that the image centroid may be estimated with relative ease and precision using a direct spatial imaging strategy, in the first step of our compound, adaptive estimation strategy~\cite{tsang2016quantum}.  

After propagating a longitudinal distance $d$, photons are detected using two cameras placed in the far field at the respective output ports of the BS, with a further spatial resolution at pixel positions $\{y_1,\dots,y_L\}$ in either camera. A detection at position $y_i$ corresponds to a resolution of the transverse momentum~(conjugate to photon transverse position) $k_i = y_iK_0/d$, where $K_0$ is the longitudinal wavenumber and $d$ is the distance to the far-field Fourier plane. The most general $L$-photon outcome of the multiphoton interference then can be described as $(X=\sum_{m}Q_m,\{y_m\});~ m\in \{1,\dots,L\}$, where $X$ is the number of photons detected in output channel $C_1$~(indicator $Q_m = 1$ iff $C_1$ is triggered), and correspondingly $L-X$ photons in $C_2$~(indicator $Q_m = 0$ iff $C_2$ is triggered), and $\{y_m\}$ is a consolidated list of the pixel positions triggered by the multiphoton detection in either camera. 

We also make the assumption that each pixel size is small enough for only one photon to be detected in a single run of the experiment, reasonably expected for $\delta k \ll 1/s$, where $\delta k$ is the range of detected transverse momenta that are integrated over inside one pixel. The inverse dependence on source separation, a feature of the Fourier plane imaging, has interesting consequences on the size of the pixels needed for effective imaging in our setup, for which
\begin{equation}\label{eq:pixelsize_condition}
    \delta y \ll \frac{d}{K_0} \frac{1}{s}.
\end{equation}
For near-IR sources with wavelength $\lambda = 2.5\,\mu m$ and table-top far-field distance of $d\sim 1\,\mathrm{m}$, Fourier plane pixel sizes must be of order $\delta y\sim 100\,\mu \mathrm{m}$ for our analysis to hold, assuming spot sizes of $\sigma_x = 100\, \mathrm{nm}$, readily available in lab as well as off the shelf~\cite{cusini2022historical,hadfield2023single,madonini2021single,incoronato2021multi,zhao2017single,cheng2023100}.

A general form for the output probabilities may be obtained starting from Eq.~(\ref{eq:lupopirandolaform})~(see Appendix \ref{app:appendix1} for details), 
\begin{widetext}
\begin{equation}\label{eq:PLform}
    P^{(L)}(X =\sum_{m}Q_m;\{k_m\}) = \prod_{m=1}^{L}|\phi(k_m)|^2~\sum_{j=0}^{L-1} \frac{ \Theta_j p_0N_s^{L-1}}{2(1+N_s(1+\delta))^{L-j-1}(1+N_s(1-\delta))^{j}}\bigg| \sum_{i=1}^L e^{i\phi_i} \xi_j(k_1,\dots,k_{i-1},k_{i+1},\dots,k_L )   \bigg|^2
\end{equation}
\end{widetext}
where $\Theta_j = \frac{(L-1-j)!j!}{X!(L-X)!}$ are factors resulting from the indistinguishability of the source and reference photons at the output BS ports, $\phi(k) = \frac{1}{\sqrt{2\pi}}\int dx e^{-ikx}\psi(x)$ are Fourier transforms of the PSF $\psi(x)$, $\xi_x$ are $L-1$-variate trigonometric functions, symmetric in their arguments, of the detected transverse momenta,
\begin{align}\label{eq:trig_fns}
    &\xi_j(k_1,\dots,k_{L-1}) =  \sum_{\substack{i_1,\dots,i_{L-1}\in \{1,\dots,L\}\\ \mathrm{all\,perms}}} \cos\left(\frac{k_{i_1}s}{2}\right)\dots
\nonumber\noindent \\
&~~~~\cos\left(\frac{k_{i_{L-1-j}}s}{2}\right)\sin\left(\frac{k_{i_{L-j}}s}{2}\right)\dots\sin\left(\frac{k_{i_{L-1}}s}{2}\right),
\end{align}
and $\phi_i$ are interferometric phase factors resulting from the propagation of $L-1$ thermal photons and the single-photon reference state across the BS~(defined in Appendix \ref{app:appendix1}). The phase factors $\phi_i$ set the signs of the trigonometric functions depending on the degree of bunching/antibunching observed at the output ports. The form of the output probabilities in Eq.~(\ref{eq:PLform}) underscores the drastic modification wrought on the photon distribution emanating from the thermal sources of arbitrary strength by interference with the non-classical state of the single photon reference, and forms the basis of the multiphoton interference sensing scheme we will henceforth construct.

We now examine explicit forms of the two-photon detection probabilities ($L = 2$) for a more transparent discussion of the interference effects, while a discussion of higher-order probabilities corresponding to $L=3,4$ is relegated to Appendix \ref{app:higherorder_prob}. We emphasise, however, that all multiphoton interference effects stemming from mixing the two-mode thermal state with a single-photon reference at the beamsplitter may be obtained in totality using Eq.~(\ref{eq:PLform}).

\subsection{Two-photon detection ($L=2$)}\label{sec:twophoton_probs}

Two-photon detections in the output ports constitute the lowest-order of non-trivial spatial interference events between the thermal source distribution and reference single-photon state that will be instrumental to the efficacy of our imaging scheme. Setting $L=2$ in Eq.~(\ref{eq:trig_fns}), we see that the two trigonometric functions that contribute to interference term in the output probabilities are $\xi_0(k) = \cos(ks/2)$ and $\xi_1(k) = \sin(ks/2)$; they yield the following form for two-photon probabilities~(recast as function of the difference~$\Delta k = k_1-k_2$ and mean of the detected momenta $\bar{K} = (k_1+k_2)/2$) :
\begin{align}\label{eq:G2_probability}
    &P^{(2)}(\bar{K},\Delta k;X) = N_s p_0^2 |\phi(\bar{K})|^2\, C(\Delta k) \,(1+N_s-\nonumber\noindent\\
    &\alpha(X)N_s\delta \cos(\bar{K}s))\left[ 1+\alpha(X)\cos\left(\frac{\Delta k}{2}s \right)  \right];X = A,B
\end{align}
where $\alpha(B) =+1$ corresponds to bunched output such that both photons are detected in the same camera, and $\alpha(X) = -1$ corresponds to anti-bunched output for photons detected in different cameras; the envelope functions have the form $C(\Delta k) = ~\frac{\mathrm{exp}[-\Delta k^2/(4\sigma_k^2)]}{\sqrt{4\pi\sigma_k^2}}$, and $|\phi(\bar{K})|^2 = ~\frac{\mathrm{exp}[-\bar{K}^2/(\sigma_k^2)]}{\sqrt{\pi\sigma_k^2}}$ for Gaussian wavepackets $\psi(x)$ with variance $\sigma_x\sigma_k = 1/2$. 

\begin{figure}
    \centering
    \includegraphics[width=0.5\textwidth]{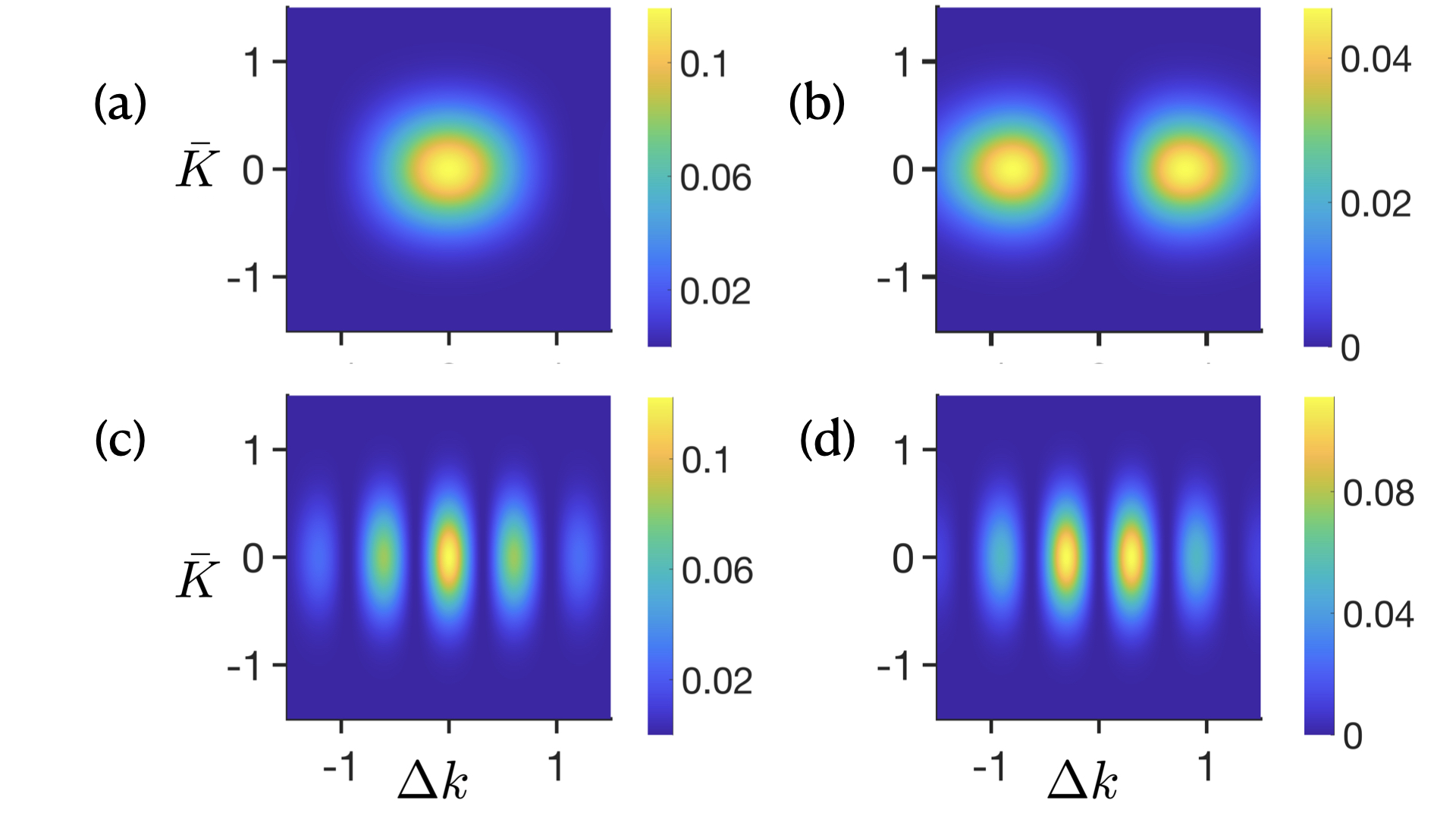}
    \caption{Two-photon detection probability $P^{(2)}(\bar{K},\Delta k;X)$ as a function of difference $\Delta k$ and mean $\bar{K}$ of the detected transverse momenta for Gaussian PSF $\psi(x)$ with $\sigma_x=1$ for $N_s=1.5$, and (a) X=B, $s=5.0\sigma_x$; (b) X=A, $s=5.0\sigma_x$; (c) X=B, $s=20.0\sigma_x$; and (d) X=A, $s=20.0\sigma_x$. With increasing separation between sources, we see that the beating in $\Delta k$ becomes comparable to the width of the transverse momentum distribution of the photons, demonstrating the sensitivity of the spatial interference pattern to source separation $s$.}
    \label{fig:2P_probability}
\end{figure}

The physical basis for the utility of multiphoton interference in estimating the separation $s$ is apparent from the beating nature of the two-photon coincident probabilities, which oscillate with a source separation-sensitive period -- $\big(4\pi/s\big)$ in $\Delta k$, and $\big(4\pi/s\big)$ in $\bar{K}$. However, interference samplings in $\Delta k$ and $\bar{K}$ are not equally visible -- while the $\Delta k$-dependent term oscillates between the numerical values $0$ and $2$, independent of the source separation 
$s$, the amplitude of oscillations in sampling variable $\bar{K}$ are damped by the PSF overlap function 
$\delta$, obscuring their usefulness in the imaging task. A more intuitive understanding of this effect may be achieved by considering for the $\bar{K}$ dependent term $f'(X;\bar{K}) = 1+N_s-\alpha(X)N_s\delta\cos(\bar{K}s)$ the visibility function 
\begin{equation}
    \mathcal{V} = \frac{f'_{\mathrm{max}} - f'_{\mathrm{min}}}{f'_{\mathrm{max}}+f'_{\mathrm{min}}} = \frac{N_s\delta}{1+N_s} <\delta,
\end{equation}
so the visibility of the $\bar{K}$ beating is maximal for $s/\sigma_x\rightarrow0$. However, such large visibilities would at the same time imply large beating periods exceeding the width of the envelope function $|\phi(\bar{K})|^2$, $2\pi/s \gg \sigma_k$, thus obscuring the $\bar{K}$ oscillations completely. On the other hand, if one were to require that the  $\bar{K}$ period is less than the full-width half-maximum~(FWHM) of the $\bar{K}$ envelope,
\begin{equation}
    2\pi/s < 2.355\sigma_k \implies \mathcal{V} <0.0285,
\end{equation}
demonstrating very low visibility for when the $\bar{K}$ oscillations fit into the spatial width of the $\bar{K}$ envelope.  This effect is also illustrated in Fig.~(\ref{fig:2P_probability}), where we note that the corresponding beating in the $\bar{K}$ make a negligible contribution, while the effect of the beating in $\Delta k$  becomes more pronounced as $s$ becomes large.

\section{Resolution Limits of Imaging Scheme}\label{sec:metrology}

So far, we have established that quantum interference with the reference single-photon state modifies photon distributions emanating from the thermal sources, producing beatings in the multiphoton coincident probabilities of detected transverse momenta that depend strongly on the source separation $s$. This raises then the following question : can we infer with enhanced precision (vis-\`{a}-vis imaging without mixing with single photon state at the BS first) the source separation $s$ by sampling from this modified interference pattern?

A separation estimator $\tilde{s}$ may be constructed from $N$ experimental runs that produce recorded outcomes $\{L^{(t)};X^{(t)},k_1^{(t)},\dots,k_L^{(t)}\}_{t=1}^{N}$, where  $L^{(t)}$ is the number of detected photons in run $t$, $\{X^{(t)},L^{(t)}-X^{(t)}\}$ is the corresponding distribution of the photons in the two output BS ports, and $\{k_i^{(t)}\}_{i=1,\dots,L^{(t)}}$ are momentum pixels for each of the recorded $L^{(t)}$ photons. The fundamental sensitivity of all estimators is bounded for our imaging setup as
\begin{equation}
    \mathrm{Var}[\tilde{s}] \geq \frac{1}{NF(s)} \geq \frac{1}{N\mathcal{Q}(s)},
\end{equation}
where the FI
\begin{align}\label{eq:fisherinfo_def1}
    F(s) &= \sum_{L=1}^{L_{\mathrm{max}}} \prod_{m=1}^L \int dk_m \frac{1}{P^{(L)}(X;\{k_m\})}\left[ \frac{\partial P^{(L)}(X;\{k_m\})}{\partial s}\right]^2 \nonumber\noindent\\
    &\equiv \sum_{L=1}^{L_{\mathrm{max}}}~F^{(L)}(s)
\end{align}
sets an upper bound to the resolution limit associated with the measurement apparatus, and $\mathcal{Q}(s)$ is the ultimate bound obtained by optimising over \emph{all} possible measurements, dubbed the quantum Fisher information~(QFI)~\cite{paris2009quantum}. Analytical forms for the QFI bound on sensitivity of separation estimators have been evaluated in \cite{lupo2016ultimate,nair2016far}. 

The quantity $F(s)$ is an important figure of merit as it sets a tight, technique-specific bound on the imaging resolution.  The component FIs $F^{(L)}(s)$ correspond to the contribution stemming from $L$-photon detection at the output, so that the sum in Eq.~(\ref{eq:fisherinfo_def1}) must run over all orders~($L_{\mathrm{max}}\rightarrow \infty$) to account for all possible multiphoton detection events. In practice, however, the likelihood of multiphoton events beyond $L_{\mathrm{max}} \gg 2 N_s+1$ drops sharply for thermal sources, meaning that it should be adequate to sum up to order $L_{\mathrm{max}} \approx f(2N_s+1), f=2-10$ to characterise the ultimate precision of the imaging scheme. This is also commensurate with finite orders of multiphoton coincidence counting that has been achieved in lab, using either spatiotemporally multiplexed camera arrays of single photon avalance detectors~(SPADs)~\cite{madonini2021single,incoronato2021multi}, or superconducting nanowire single photon detectors~(SNSPDs)~\cite{zhao2017single,cheng2023100} where spatially-resolved coincidences up to photon number $4$~\cite{incoronato2021multi} and $15$~\cite{zhao2017single} respectively have been detected.

In the remainder of this section, we will examine the characteristics of quantum imaging corresponding to different multiphoton orders that can be implemented in our scheme by studying the behaviours of various orders of $F^{(L)}(s)$ as functions of source separation $s$ and brightness $N_s$. This will be done in three partitions -- first, we will examine precision limits in terms of FI corresponding to all orders of photon detection for sub-Rayleigh separations ($s/\sigma_x\ll1$) where we will establish the available quantum superresolution of our imaging scheme; second, we will examine precision limits of our scheme for large separations ($s/\sigma_x\gg1$) where a finite $F(s)$ demonstrates the large spatial range of the scheme's efficacy; and finally we will present numerical results for $F^{(L)}$ for intermediate values of separation that can not be evaluated analytically. All numerical results will be restricted to order $L\leq 7$ to aid transparent qualitative discussion of imaging scheme characteristics. Finally, a detailed discussion of two-photon sampling precisions for arbitrary separations $s/\sigma_x$ and $N_s$ for which analytical results are accessible is relegated to Appendix \ref{sec:2PFIs}. 

\subsection{Sub-Rayleigh Separations $s/\sigma_x\ll 1$}\label{section:subrayleigh_seps}

\begin{figure}
    \centering   \includegraphics[width=0.45\textwidth]{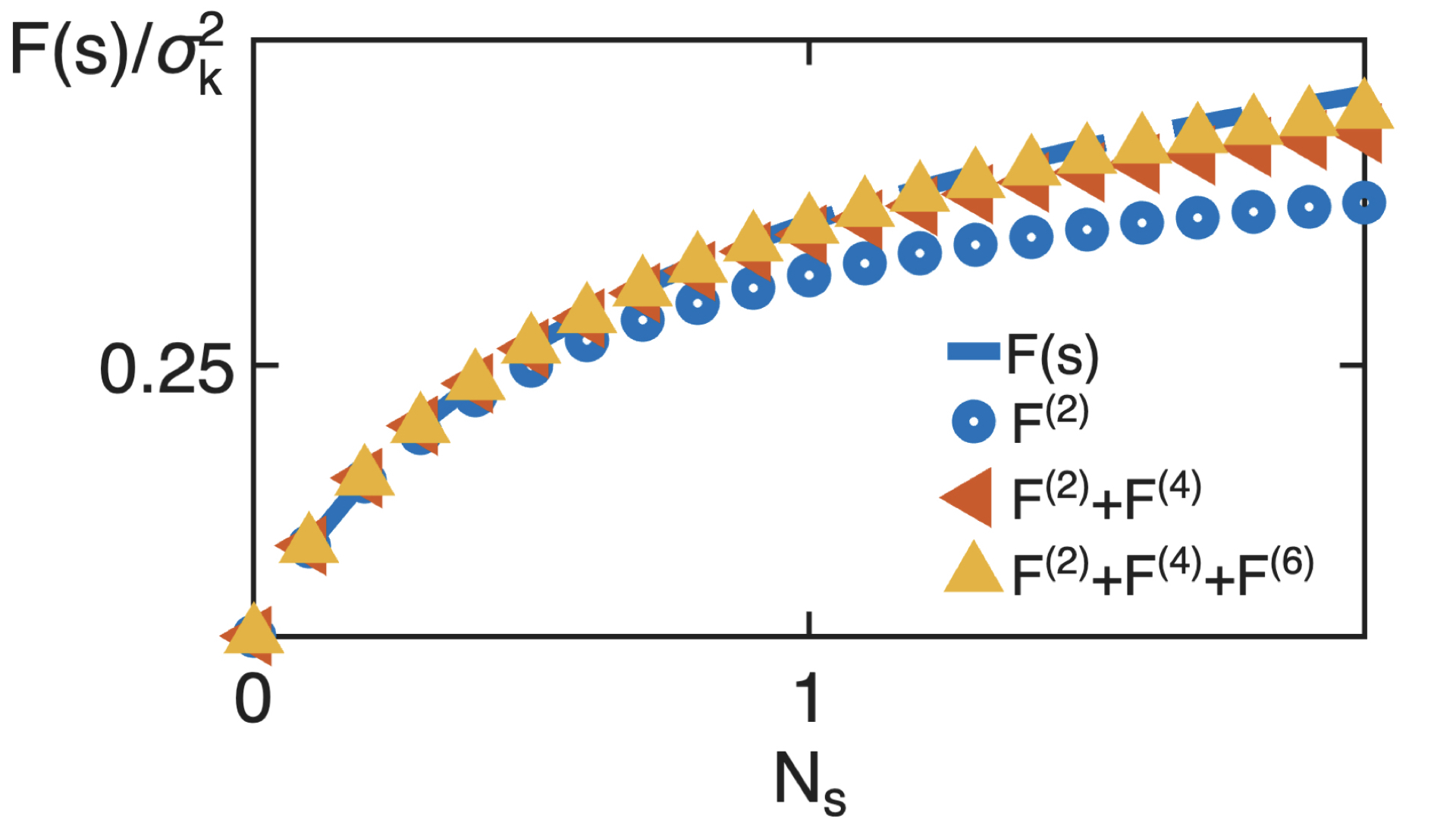}
    \caption{Sub-Rayleigh FI $F(s/\sigma_x\rightarrow 0)$~(blue dashed line) as a function of source brightness $N_s$, and $\sigma_x=1.0$, from the closed expression in Eq.~(\ref{eq:FIsubrayleigh_allorders}). Also displayed are sub-Rayleigh FIs for detection up to multiphoton coincidence orders $L=2$~(blue circles), $L=4$~(orange-rotated triangles), and $L=6$~(yellow triangles), demonstrating that for $N_s\sim 1$, lower order photon detection is sufficient to saturate almost all of the available imaging precision.}
    \label{fig:sub_rayleighplot}
\end{figure}

In order to investigate the sub-Rayleigh performance of our scheme, the coincident probabilities in Eq.~(\ref{eq:PLform}) may be expanded as a power series around $s/\sigma_x\rightarrow 0$ by leveraging the Fock decomposition of the transformed uncorrelated thermal sources in Eq.~(\ref{eq:lupopirandolaform}). Following from the general relation $\partial \rho_d/\partial s \propto s \sigma$~(for some Hermitian operator $\sigma$ and $\rho_d$ defined in Eq.~(\ref{eq:rhod})) established in \cite{oh2021quantum}, we have for detection probabilities,
\begin{equation}
    \frac{\partial}{\partial s}P^{(L)} (X;\{k_m\})= s~\mathcal{F}(X;\{k_m\});~ \mathcal{F} \in \mathds {R}.
\end{equation}
It is then easy to deduce that any multiphoton distribution $P^{(L)}$ whose respective FI does not vanish for sub-Rayleigh separations $F^{(L)}[s/\sigma_x \rightarrow 0] >0$ --- dubbed ``superresolution" in quantum imaging as it corresponds to imaging beyond the limits of the Rayleigh criterion --- must scale \emph{exactly} as $O(s^2)$. This rather strong requirement has two consequences : a) only even-order photon detections contribute to the sub-Rayleigh FI and estimator precisions, and b) only balanced antibunching events for even-order detections~(so equal number of photons are detected in each output BS port) contribute any information in the sub-Rayleigh limit, and all other events contribute no information.

More concretely, in the the sub-Rayleigh limit the leading term in $P^{(L)}$ stems from the coincident probabilities for detecting $L=P$ photons in each output BS port at pixel positions $\{k_i\}_{i=1,\dots,2P}$, which takes the explicit form~(for details, see Appendix \ref{app:subRayleigh_app})
\begin{align}\label{eq:PLform_subrayleigh}
    &P^{(L=2P)}(X=P;k_1,\dots,k_{2K})=\frac{(2P-2)!}{P!P!}\frac{N_s^{2P-1}}{(1+2N_s)^{2P-1}}\nonumber\noindent\\
    & \prod _{m=1}^{2P} |\phi(k_m)|^2 (k_1+\dots k_{P}-k_{P+1}-\dots -k_{2P})^2~\frac{s^2}{4} + O(s^4),
\end{align}
in orders of $s$, yielding the following form for limiting FIs from the $2P$-correlator of the multiphoton interference:
\begin{equation}\label{eq:fisher_subrayleigh}
    \frac{F^{(2P)}(s/\sigma_x \rightarrow 0)}{\sigma_k^2} =  \frac{\binom{2P}{P}}{2(2P-1)}\left( \frac{N_s}{1+2N_s} \right)^{2P-1}.
\end{equation}
What this tells us then is striking : all even-order correlators of multiphoton interference can be used to build estimators of separation $s$ whose precisions do not vanish for the sub-Rayleigh regime. This is a direct consequence of the extended HOM effect, of which Eq.~(\ref{eq:PLform_subrayleigh}) is a manifestation particular to our setup. However, the metrological analysis establishes here for the first time that the extended HOM yields superresolution in incoherent imaging of strong thermal sources, just as two-photon HOM yields superresolution in imaging weak thermal sources~\cite{muratore2025superresolution}. 

Next, note that $F^{2P} > F^{2P'}$ iff $P<P'$, meaning that lower order correlators are always more informationally dense, a direct consequence of the thermal mode structure of the image state such that the probability of recording multiphoton events decays exponentially with the number of photons detected. Even so, we note that higher order correlators become more informative as $N_s$ increases (even though $F^{2P} > F^{2P'}$ holds for all $N_s$) , a behaviour that was already noted in Figure \ref{fig:FIplotseparation}. The fundamental bound on the sensitivity possible in our multiphoton interference scheme in the sub-Rayleigh limit can be obtained by summing over all even-order correlators,
\begin{align}\label{eq:FIsubrayleigh_allorders}
    \frac{F(s/\sigma_x \rightarrow 0)}{\sigma_k^2}  &=  \sum_{P=1}^{\infty}\frac{F^{(2P)}(s/\sigma_x \rightarrow 0)}{\sigma_k^2}\nonumber\noindent\\
    &= \frac{1+2N_s-(1+4N_s)^{1/2}}{2N_s}.
\end{align}
Figure \ref{fig:sub_rayleighplot} illustrates the relative contributions of the two-photon FI $F^{(2)}$ as a ratio of the total imaging FI $F(s\rightarrow 0)$, where we see that for faint sources $N_s\ll 1$, two-photon FIs retrieve all of the information, whereas for large $N_s$, higher orders are at least as important, tending to the value of
\begin{equation}
    \lim_{N_s\rightarrow \infty} \frac{F^{(2)}(s\rightarrow 0)}{F(s\rightarrow 0)} \rightarrow \frac{1}{2},
\end{equation}
which can be obtained from the analytical form of Eqs.\, (\ref{eq:fisher_subrayleigh})  and (\ref{eq:FIsubrayleigh_allorders}). This limiting form of the imaging FI is plotted as a function of $N_s$ in Figure \ref{fig:sub_rayleighplot}, where it can be seen that for $N_s\sim 1$, FIs $F^{(L)}$ up to order $6$ are sufficient to approximately saturate $F(s)$.

\subsection{Large Separations $s/\sigma_x \gg 1$}
For asymptotically large separations between the incoherent sources, we have $\lim_{s/\sigma_x \rightarrow\infty}\delta =0$, and the following form for the $L$-photon coincident probabilities:
\begin{equation}
    P^{(L)} = \frac{N_s^{L-1}}{(1+N_s)^{L+1}}\prod _{m=1}^L |\phi(k_m)|^2 \Upsilon_{k_1,\dots,k_L}(s),
\end{equation}
where 
\begin{equation}
\Upsilon_{k_1,\dots,k_L}(s) = \sum_{j=0}^{L-1} \frac{\Theta_j}{2} \bigg| \sum_{i=1}^{L-1} p_i \xi_j(k_1,\dots,k_{i-1},k_{i+1},\dots,k_{L})  \bigg|^2
\end{equation}
is an oscillating function of the separation $s$. This then yields the following form for the FIs for $s/\sigma_x\gg1$:
\begin{equation}\label{eq:largeS_multiphoton}
    F^{(L)}(s) = \frac{N_s^{L-1}}{(1+N_s)^{L+1}} \left\langle \frac{[\partial_s  \Upsilon_{k_1,\dots,k_L}(s)]^2}{\Upsilon_{k_1,\dots,k_L}(s)}  \right\rangle_{k_1,\dots,k_L}
\end{equation}
where $\langle \dots\rangle_{k_m}$ correspond to Gaussian mean in the transverse momentum distribution $|\phi(k_m)|^2$. Although closed expressions for arbitrary $L$ photons are hard to obtain in this case~(see Appendix \ref{app:2p_largesep} for analytical forms of $F^{(2)}(s/\sigma_x\gg1)$), the form in Eq.~(\ref{eq:largeS_multiphoton}) allows us to make a few useful observations. 

First, we note that  $F^{(L)}(s)$ for $s/\sigma_x\gg1$ scales as $N_s^{L-1}/(1+N_s)^{L+1}$, in contrast to the sub-Rayleigh scaling $N_s^{L-1}/(1+2N_s)^{L-1}$ for even $L$. For source brightness up to $N_s\sim 1$, this also means that $F^{(L)}(s/\sigma_x\gg 1) = O(N_s)$, matching the optimal quantum ordering of $\mathcal{Q}(s)$  More directly, we also see that the $N_s$ dependence of the FI matches the probability of the thermal sources collectively emitting $L-1$ photons, reflecting the fact that the two-mode state that is imaged for $s/\sigma_x\gg 1$ is uncorrelated in the spatial bases. 

Second, the form in Eq.~(\ref{eq:largeS_multiphoton}) also implies that the $L$-photon FIs $F^{(L)}(s/\sigma_x\gg 1)$ are maximised with respect to $N_s$ for $N^{\mathrm{L}}_{s,{\mathrm{max}}} = (L-1)/2$, as also apparent from Figure \ref{fig:multiphotonscaling_FIs} (c) for $L\in \{2,\dots,7\}$. This has practical consequences for our imaging scheme, as it means that for sources with brightness $N_s\sim 1$, most of the contributions come from lower order photon events corresponding to $L\in \{2,\dots,7\}$.

\subsection{Intermediate Separations}

 \begin{figure}
       \centering
       \includegraphics[clip,width=\columnwidth]{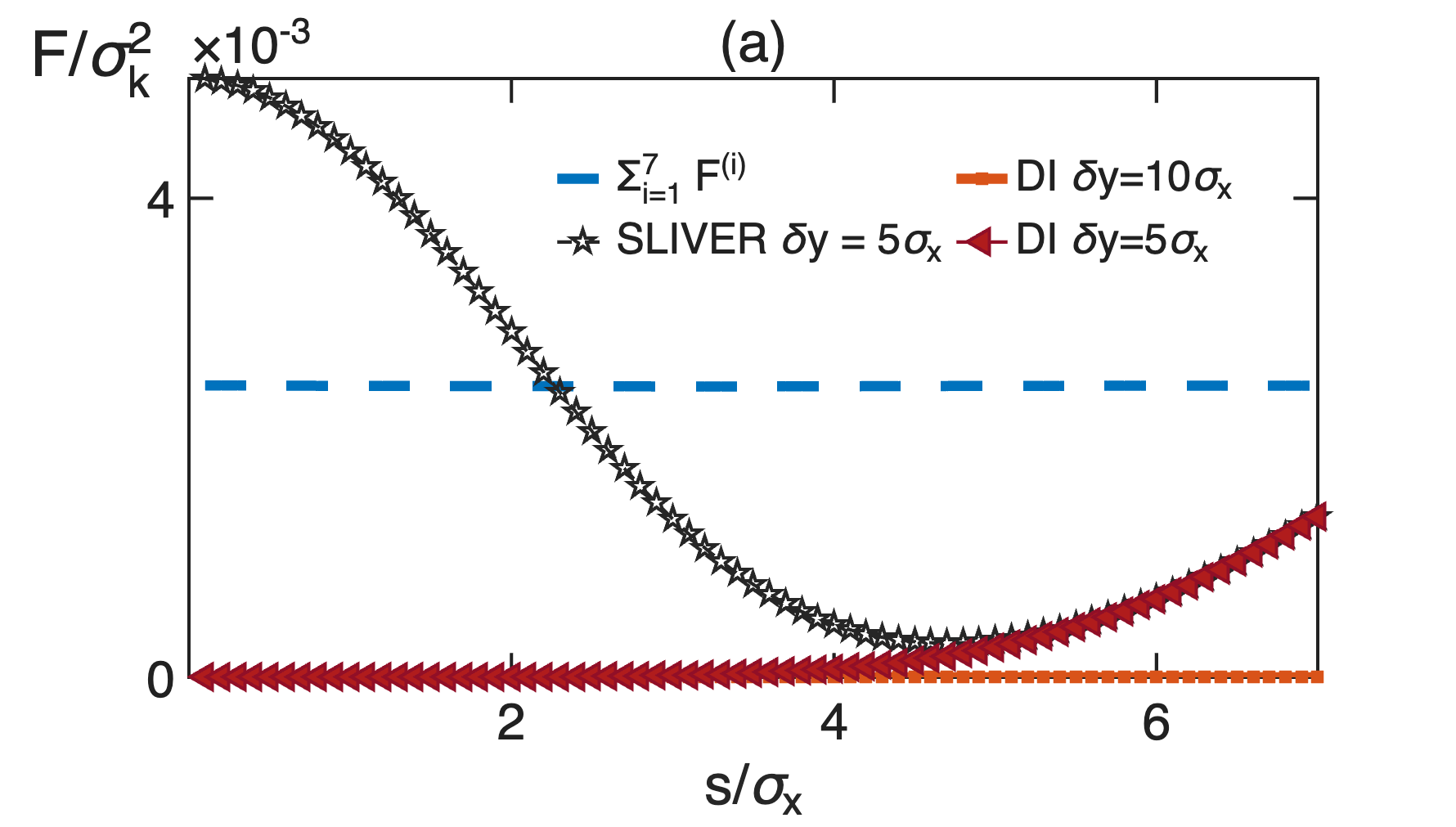}

       \includegraphics[clip,width=\columnwidth]{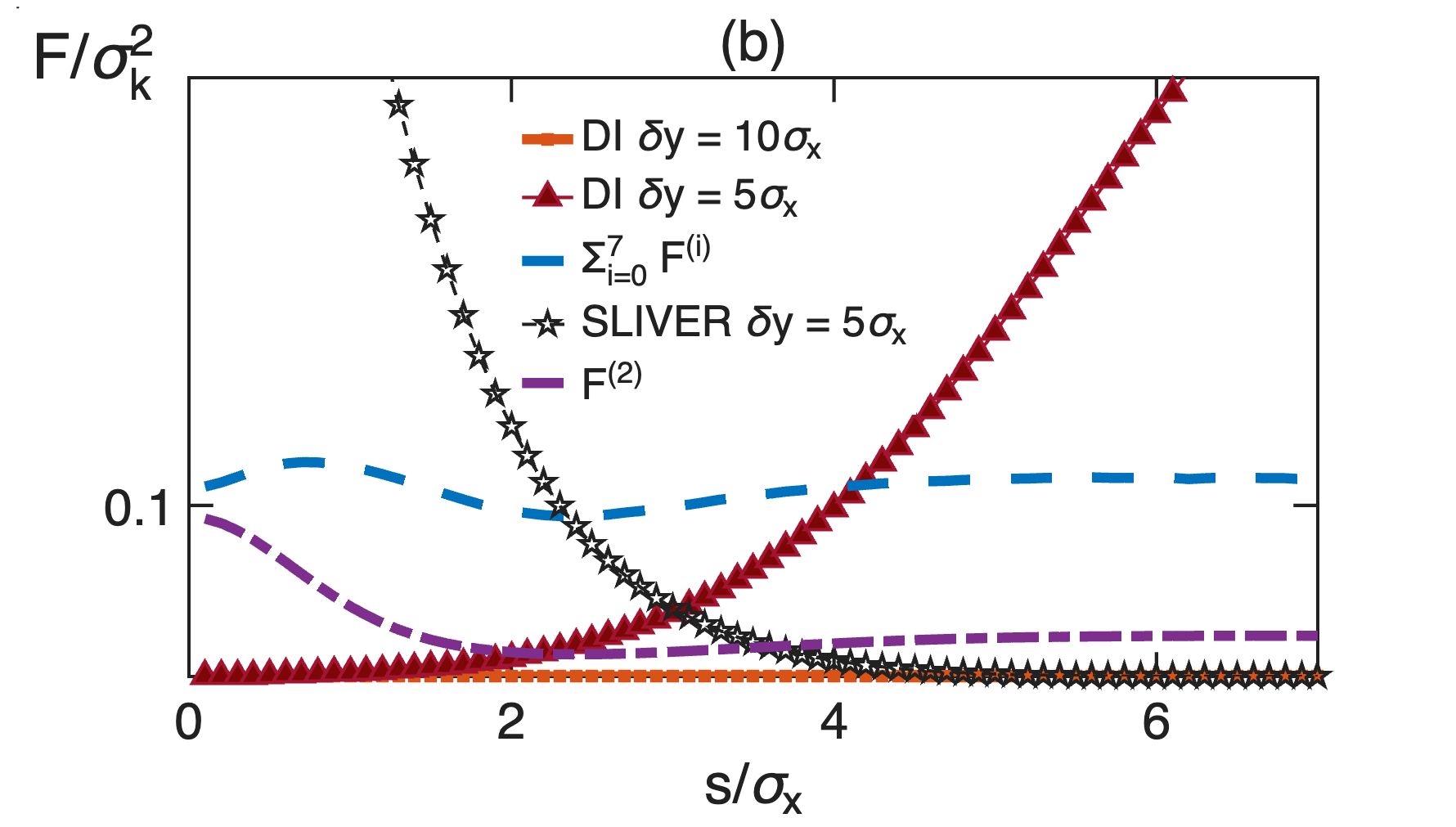}
       
       \caption{Imaging FI $\sum_{i=1}^L F^{(i)}(s) \sim F(s)$~(blue dashed line) for coincident multiphoton sampling up to $i\leq L=7$ as a function of source separation $s/\sigma_x$, for (a) $N_s=0.01$~(feeble sources), and (b) $N_s=1.5$~(bright sources). For a meaningful comparative, we have also superposed lower bounds~\cite{nair2016far} for pixelated direct imaging FI for pixel sizes $\delta y=5\sigma_x$~(maroon solid line-triangle), and $\delta y=10\sigma_x$~(orange solid line-square); the lower bounds for pix-SLIVER FI~(black dotted line-pentagrams) for pixel size $\delta x = 5\sigma_x$; and FI of only two-photon events $F^{(2)}$ (magenta dot-dashed) in (b). All multidimensional numerical integrations were performed using routines from the Cuba library~\cite{hahn2005cuba}.  }
    \label{fig:FIplotseparation}
   \end{figure}

 \begin{figure*}
       \centering
       \includegraphics[width=0.32\linewidth]{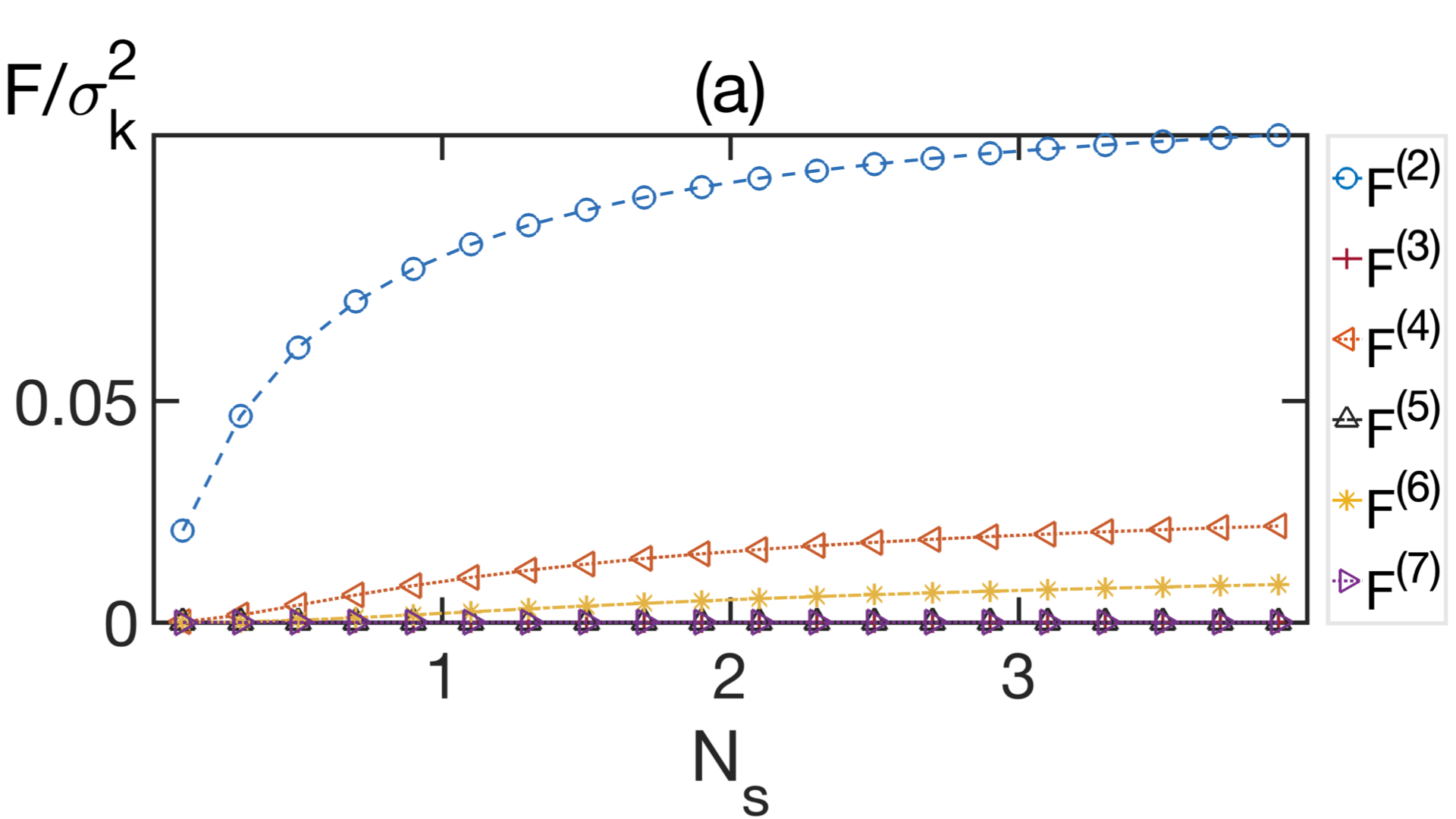}
       \includegraphics[width=0.32\linewidth]{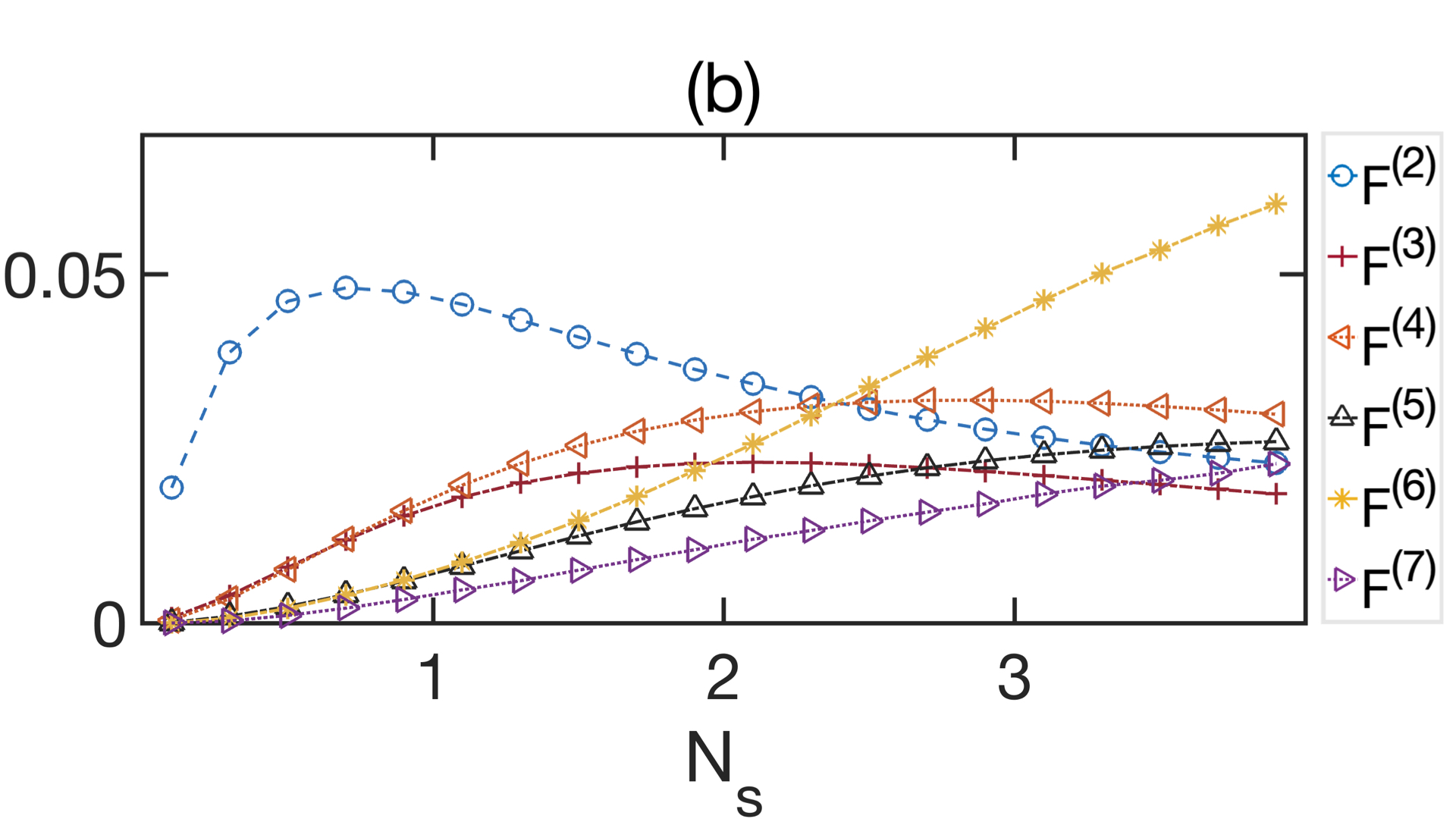}
       \includegraphics[width=0.32\linewidth]{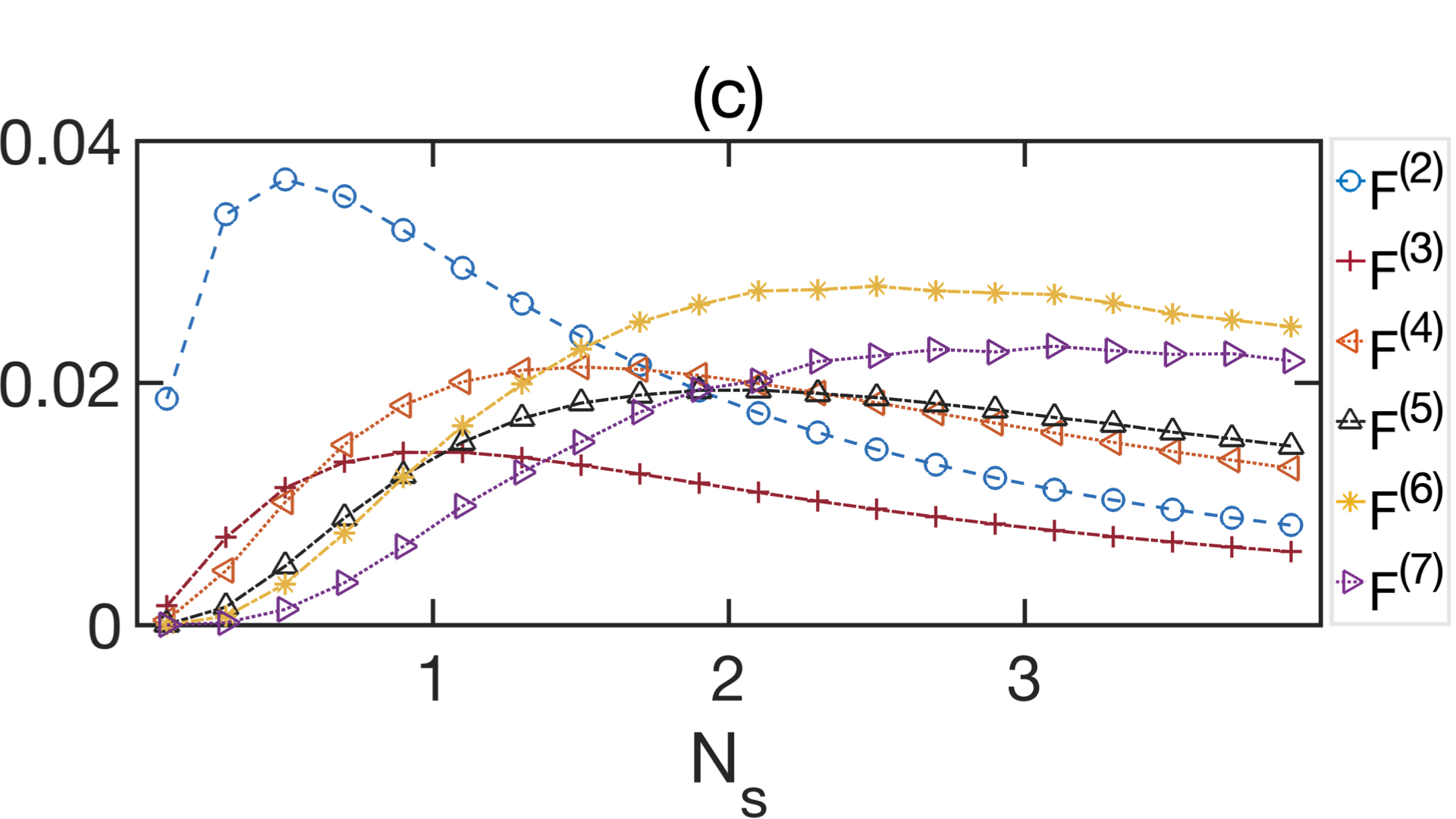}
       \caption{Comparative analysis of contributions to total multiphoton FI $F(s)$ for the imaging scheme from photon orders corresponding to $L=\{2,\dots,7\}$, as a function of source brightness $N_s$, and for (a) $s/\sigma_x = 0.01$, (b) $s/\sigma_x = 1.0$, and (c) $s/\sigma_x = 7.0$. For sub-Rayleigh separations, the analytically established ordering $F^{(2)}>F^{(3)}>F^{(4)}>\dots$ holds, but for intermediate and large separations, higher order FIs become increasingly important, and may even contribute more information than lower order FIs. All multidimensional numerical integrations were performed using the Cuba library~\cite{hahn2005cuba}.}
       \label{fig:multiphotonscaling_FIs}
   \end{figure*}

FIs from multiphoton coincidence detection of any order $L$ at the output BS ports may be evaluated numerically using the explicit form of output probabilities in Eq.~(\ref{eq:PLform}). For typological values of $N_s$ characterising feeble~($N_s=0.01$) and relatively bright sources~($N_s=1.5$), the FI associated with multiphoton events up to $L=7$ are displayed in Figure \ref{fig:FIplotseparation}. The choice $L\leq 7$ is commensurate with the chosen values of brightness for which higher-order multiphoton events beyond $L=7$ are rare, and contribute almost no FI. This then allows us to numerically approximate the total imaging FI as $F(s) \approx \sum_{i=1}^7 F^{(i)}(s)$. 

For comparison, we also plot lower bounds to FIs of competing imaging techniques of pixelated DI~(assuming pixel sizes $\delta y=5,10\sigma_x$) and pix-SLIVER~(assuming pixel size $\delta y=5\sigma_x$). Our choice for $\delta y$ reflects the smallest pixels available in lab for spot size $\sigma_x  = 0.1\,\mu$m~\cite{hu2015superconducting,cusini2022historical}; the FI bounds for DI and SLIVER
then correspond to maximum achievable precision with practically available pixel sizes. A fair comparison between the pixelated FI bounds for DI and SLIVER and non-pixelated $F(s)$ is reasonable because
the Fourier plane imaging in our multiphoton scheme
means pixels as large as $100\sigma_x$ will still respect the condition
set in Eq.~(\ref{eq:pixelsize_condition}) and saturate non-pixelated $F(s)$.

Our numerical results demonstrate, first and foremost, that for intensities exceeding the weak source model, detection of multiphoton coincidences beyond two-photon events significantly enhance precision. For weak thermal sources $N_s\ll 1$ in Figure \ref{fig:FIplotseparation}(a), two-photon detection retrieves almost all of the information for all separations $s$~\cite{muratore2025superresolution} as the probability of the sources collectively emitting more than a single photon is quite small. In contrast, for bright thermal sources $N_s\sim 1$ in Figure \ref{fig:FIplotseparation}(b), we see that there is a widening gap between the two-photon FI $F^{(2)}$ and full FI $F(s) \approx \sum_{i=1}^7 F^{(i)}$. This is especially true for moderate to large values of source separation $s$ where the total multiphoton $F(s)$~(and hence correspondingly lower bound on all estimator precisions) can be several times larger than just the two-photon FI $F^{(2)}$. This enhancement afforded by detection of multiphoton events extends the region of supremacy over pixelated DI --- for $N_s=1.5$ and DI pixel sizes $\delta y=5\sigma_x$ in Figure \ref{fig:FIplotseparation}(b), while two-photon detection yields more precise estimators than direct imaging for $s/\sigma_x \lessapprox 2.0$, detecting up to $L=7$ photons approximately doubles this interval of quantum advantage to $s/\sigma_x \lessapprox 4.0$. We also observe that for DI pixel size $\delta y=10\sigma_x$, both $L=2$ and $7$ photon detection schemes have uniform advantage over all separations $s$. 

Furthermore, the transverse momenta resolution using coarse pixels in the Fourier plane further affords our scheme advantage over pix-SLIVER in the regime of moderate to large separations. In both Figures \ref{fig:FIplotseparation} (a) and (b), we see that the imaging FI from our scheme exceeds pix-SLIVER FI bound beyond $s/\sigma_x\gtrsim 2.5$, while still offering finite resolution in the sub-Rayleigh regime where the imaging FI is a fraction of the total QFI.
 
What we have demonstrated then is that by the resolving multiphoton events of around the same order as the source brightness $N_s$ in the Fourier plane, our scheme offers robust, near constant FI over a wide range of separation $s$, for \emph{both} feeble and relatively bright sources. This is in stark contrast with peer imaging schemes of pix-DI and pix-SLIVER that offer very little resolution respectively for small and large separations $s$, even when we allow for the use of smallest pixel sizes technologically available. This has interesting consequences for applications such as single-particle tracking~(SPT) in label microscopies and spectroscopies~\cite{manzo2015review,moerner2015single} where dynamically changing separations between fluorophores may indicate conformational changes or the appearance of rare transition states. These fluctuating inter-source distances may be tracked much more accurately with our multiphoton scheme than either pix-DI or pix-SLIVER.

On balance, multiphoton detection beyond $L>2$ is challenging for bright sources that are usually encountered in the infrared~(IR) regime where background noise is large. However, we note that current experimental landscape around pixelated arrays~\cite{madonini2021single,incoronato2021multi,cheng2023100,zhao2017single} made of SPAD and especially SNSPD single-photon detectors is commensurate with typical source brightnesses encountered in lab of $N_s \sim 1$, where our imaging scheme is near-optimal for \emph{all} separations with up to $L=7$ detection events.

Finally, we also note that the hierarchy of magnitudes of contributions $F^{(L)}$ to the total FI $F(s)$ critically depends on the separation $s$. This is illustrated in Figure \ref{fig:multiphotonscaling_FIs} (a) where we see that $F^{(2)}>F^{(4)}>F^{(6)}$ for all $N_s$ for sub-Rayleigh separations~(already analytically established in Section \ref{section:subrayleigh_seps}). 

In contrast, two-photon detections become relatively less informative as separation $s$ increases~(see Figures \ref{fig:multiphotonscaling_FIs} (b) and (c)) where there is an emergence of critical $N_s$ beyond which $F^{(7)}<F^{(6)}<F^{(5)}<F^{(4)}<F^{(3)}<F^{(2)}$ holds. The onset of this reverse ordering occurs for increasingly small $N_s$ as separation $s$ increases -- this is again a manifestation of the FIs $F^{(L)}$ scaling as the probability of corresponding emission of $\frac{L-1}{2}$ photons from the source in the limit of $s/\sigma_x \rightarrow \infty$, a behaviour we can readily observe approximately from Figure \ref{fig:multiphotonscaling_FIs} (c). This also reinforces our earlier conclusion that multiphoton events beyond $L=2$ are increasingly informative for bright sources, in the regime of moderate to large separations.

\subsection{Resolving versus Non-Resolving Detectors}
 \begin{figure*}
       \centering
       \includegraphics[width=0.32\linewidth]{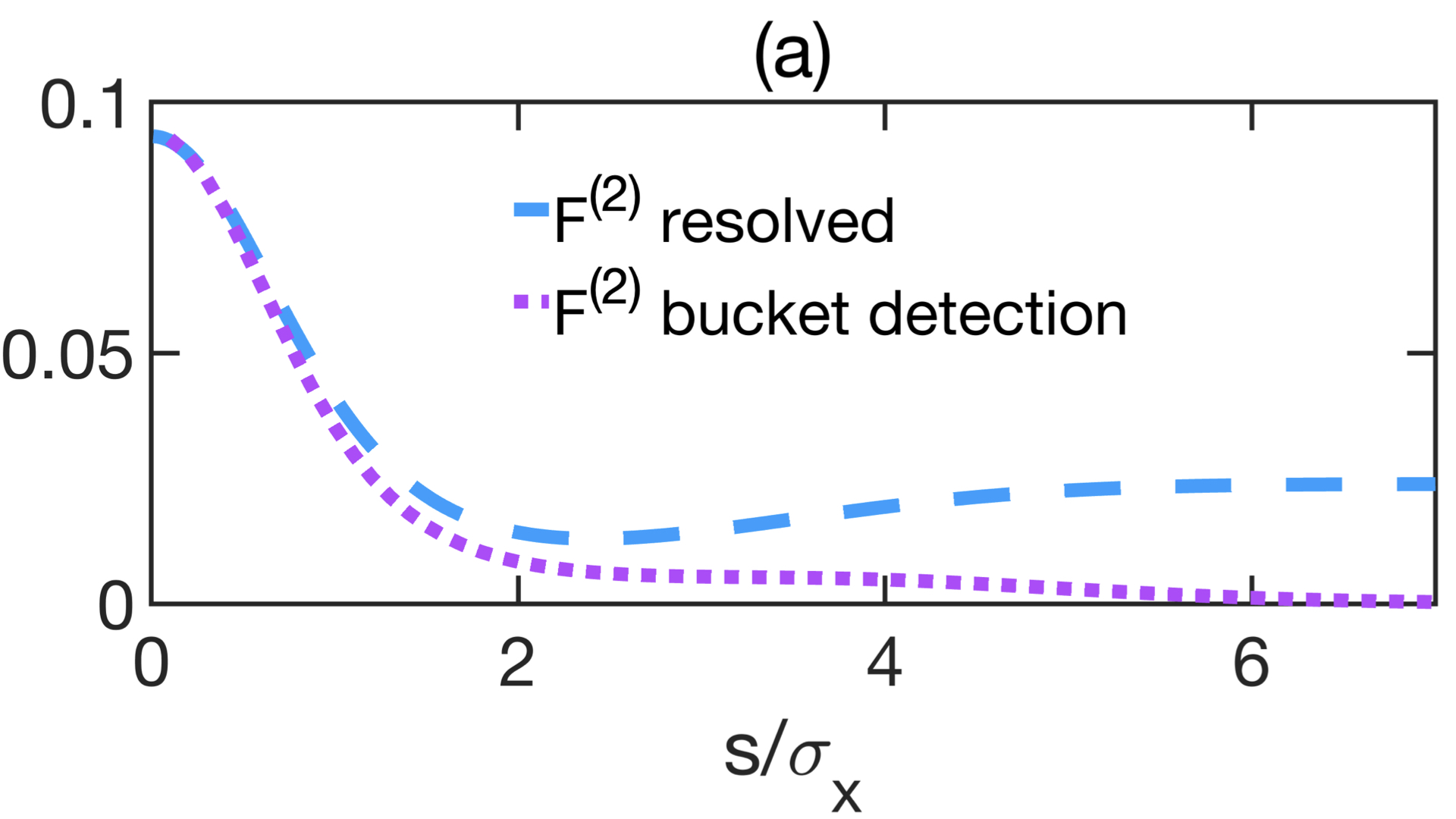}
       \includegraphics[width=0.32\linewidth]{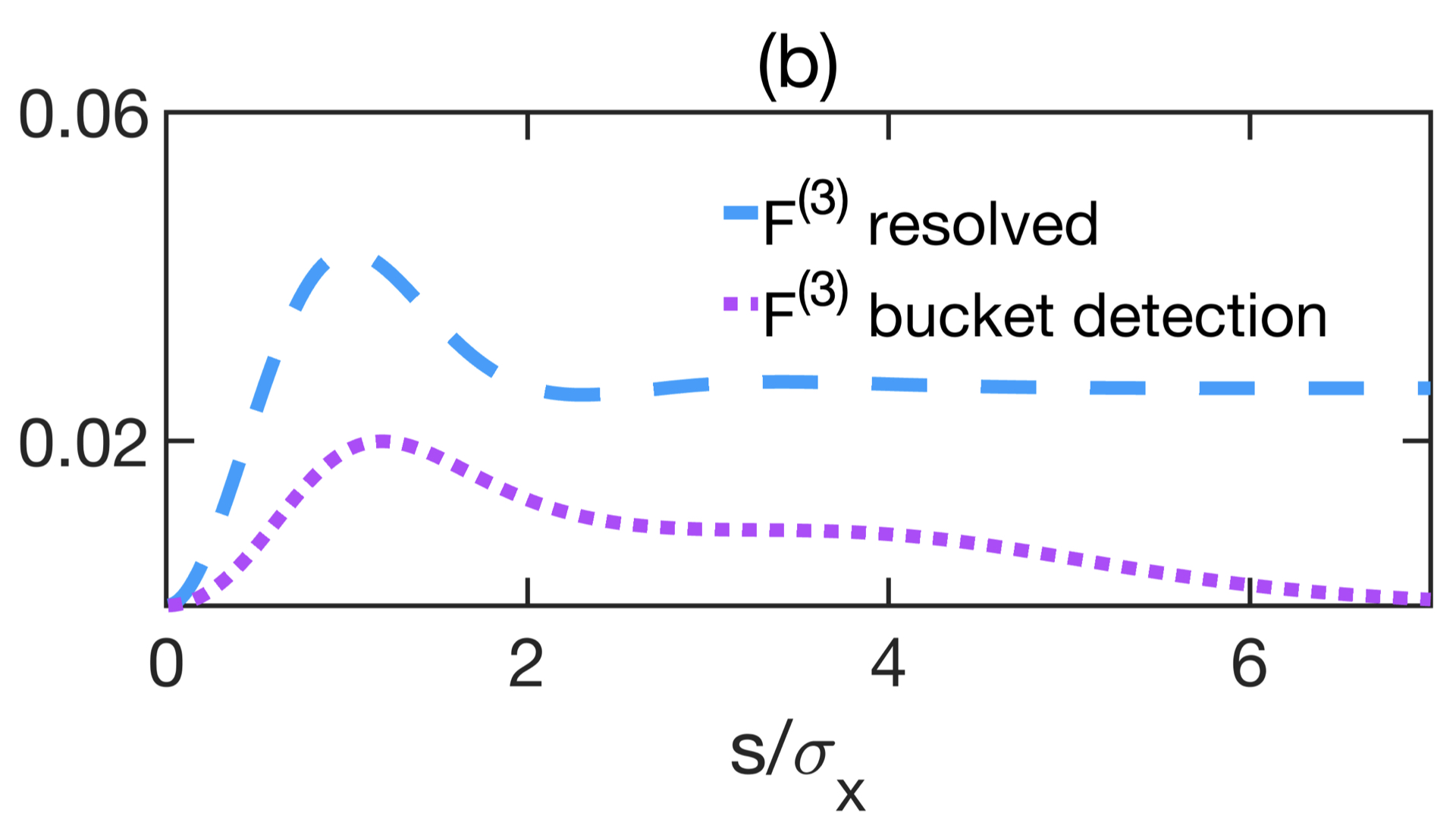}
       \includegraphics[width=0.32\linewidth]{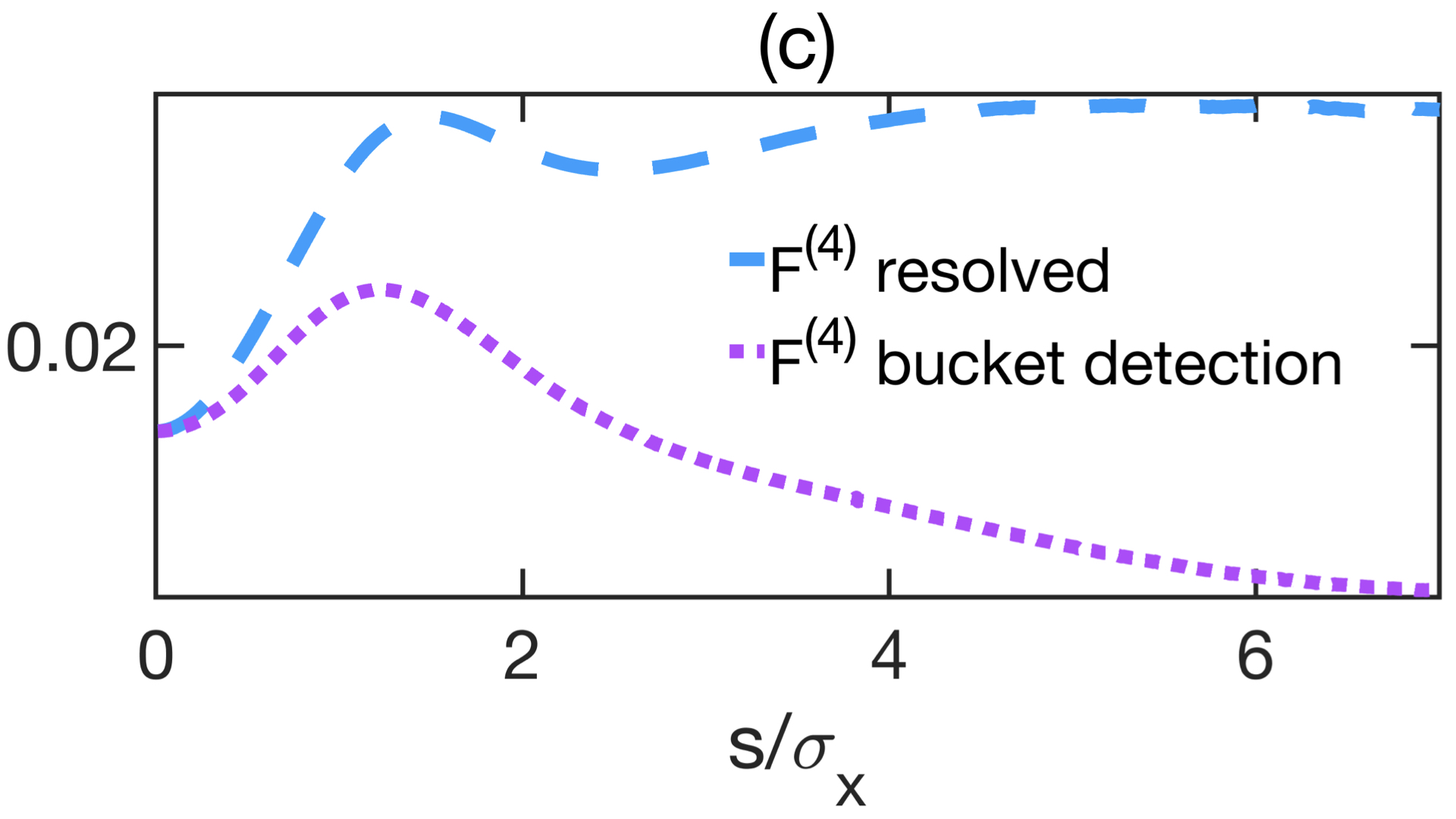}
       
       \caption{FIs for transverse momenta resolving as well as bucket detections for photon orders of (a) $L=2$, (b) $L=3$, and (c) $L=4$ as a function of separation. For small separations, bucket detection retrieves all of the FI, whereas only transverse momentum resolution yields estimator precisions for large separations $s/\sigma_x$. All multidimensional numerical integrations were performed using Cuba library routines~\cite{hahn2005cuba}.~($N_s=1.5$) }
       \label{fig:resolvedvnonresolved}
   \end{figure*} 
The multiphoton interference scheme bears an additional step vis-\`{a}-vis typical HOM interferometers in its resolution of inner variable conjugate to the parameter being estimated, which in this case is the source separation $s$. In fact, the transverse momentum resolution effected by pixelated arrays in the far field renders the detectors completely blind to the source separation~\cite{triggiani2024estimation}, yielding the multiphoton beating in transverse momenta variables. These beatings, with periods proportional to $1/s$, are directly responsible to finite FI even when the spatial PSFs corresponding to the two sources have no overlap. This is illustrated in Figure \ref{fig:resolvedvnonresolved} for photon orders up to $4$, where we see that for $s/\sigma_x>1$, bucket detectors that do not resolve the transverse momenta yield no information, whereas resolved detection yields a finite amount.

In contrast, for sub-Rayleigh separations, superresolution of the multiphoton scheme is independent of the transverse momentum resolution afforded by the far-field pixels, but is rather a property of the extended HOM effect at the beamsplitter that creates the ``rare" balanced anti-bunched events whose vanishing probabilities for $s/\sigma_x\rightarrow0$ yield large signal-to-noise~(SNR) for such events in the sub-Rayleigh regime. This can be quantified by noting that the transverse momenta-integrated probability of balanced anti-bunching~($X=P$) is 
\begin{equation}
    P^{(L=2P)}(X=P) = \frac{\binom{2P}{P}}{2(2P-1)} \left( \frac{N_s}{1+2N_s}\right)^{2P-1}~\frac{s^2}{4}
\end{equation}
yielding the same FI in the limit $s/\sigma_x\rightarrow0$ as does $ P^{(L=2P)}(X=P;\{k_m\})$. However, it is important to note that this simplified measurement strategy still requires some manner of number resolution~(except for $L=2$) as one must still be able to discern the number of photons detected in each output port, though it does not have to be necessarily in the far field. This analytical insight is substantiated in Figure \ref{fig:resolvedvnonresolved} for orders up to $L=4$, where we see that bucket detectors are sufficient to obtain superresolution.

\section{Conclusions and Outlook}\label{sec:conclusions}
We have presented here a sensing scheme to resolve separations between incoherent bunched sources of arbitrary brightness, using the quantum interference effect of eHOM between the multiphoton field emanating from these incoherent sources and a single reference photon when spatially resolved measurements are employed in the far field. We have established quantum superresolution in the sub-diffraction limit for arbitrary even-order photon coincidences, a drastic advantage over any classical imaging strategy. In the well-established setup of the HOM interferometer, we also demonstrate supremacy over direct imaging as well as two-photon coincidence measurements over a wider range of source separations in terms of fundamental sensitivities quantified by FIs, showing that our proposed technique provides a practical and experimentally robust method for novel applications such as SPT of live cells. 

By harnessing multiphoton interference phenomena, our imaging strategy provides a simplified route to quantum imaging beyond the Rayleigh limit. As the superresolution in the sub-diffraction limit is vulnerable to effects of detection noise, the elimination of crosstalk and alignment errors is an important step for imaging bright sources. With the technological advancement in single photon detector array cameras over the last decade, especially SNSPD cameras for imaging bright IR sources where evanescent noise floor is substantial, the ability in particular to resolve coincident multiphoton detection events beyond two photons at a time presents a straightforward way to realising the quantum advantage in imaging super-Poissonian sources in nanoscopic applications, which our scheme readily extracts.

Finally, the finite imaging precision obtained in our scheme for finite, practically accessible pixel sizes for both sub- and super-Rayleigh separations is a unique feature -- pix SLIVER exhibits superresolution in the sub-Rayleigh, but bad precision for large separations; on the other hand, direct imaging is quite handy for large separations but offers no superresolution. Using a large number of coarse inexpensive pixels in a simple optical setup with no crosstalk, our scheme presents an experimentally viable alternative to quantum superresolution imaging of biological, chemical, and  stellar targets including and beyond the visible spectrum.

\begin{acknowledgments}

This project was partially supported by the Air Force Office of Scientific Research under award number FA8655-23-1- 7046. A.K. acknowledges helpful discussions with Francesco Albarelli, Gerardo Adesso, Samanta Piano, George Brumpton. D.T. acknowledges the Italian Space Agency (ASI, Agenzia Spaziale Italiana) through the project Subdiffraction Quantum Imaging (SQI) n. 2023-13-HH.0. 

\end{acknowledgments}

\begin{widetext}

\appendix

\section{Coincident probability $P^{(L)}(\{k_i,C_i\})$ in Eq.~(\ref{eq:PLform})}\label{app:appendix1}
The detection operator for the transverse momenta-resolved ``which-detector" measurement at output camera $Q = 1 \,\,(\mathrm{for}\,\, C_1)\,\,\mathrm{or}\,\, 0\,\,(\mathrm{for}\,\, C_2)$ placed a transverse distance $d$ is given  as~\cite{triggiani2024estimation,muratore2025superresolution}
\begin{equation}\label{eq:detection_op}
    E_{Q}^+(k) = \frac{1}{\sqrt{2\pi}}\sum_{S=0,1}\int dx\, g(x,S;k,Q)\, a_{S}(x); ~~g(x,S;k,Q) = \frac{e^{i\Phi(S,Q)}}{\sqrt{2}}~e^{-iK_0d}e^{-ikx},
\end{equation}
where $K_0$ is the momentum along the propagation direction, $k = yK_0/d$ is the transverse photonic momentum, and 
\begin{equation}
    \Phi(S,Q) = \begin{cases}
        0, ~S=Q \\
        \pi/2,~ S \neq Q
    \end{cases}
\end{equation}
are interferometric phases for a balanced BS acquired as the photons propagate from input port $S=0,1$ to the output ports $Q=1$\,(for $C_1$) and $Q=0$\,(for $C_2$). In terms of momentum space operators, $a_S(k) = \frac{1}{\sqrt{2\pi}}~\int dx ~e^{-ikx}a_S(x)$, the electric field at the camera $Q$ can be re-expressed as 
\begin{equation}
    E_{Q}^{+}(k) = \sum_{S=0,1} E_{SQ}^{+}(k);~~E_{SQ}^{+}(k) = \frac{e^{i\Phi(S,Q)}}{\sqrt{2}}~a_S(k).
\end{equation}
The most general detection outcome for the setup $\{n(k_i,Q_i)\}$ is a number-resolved detection of $n$ photons at pixel position $y_i = k_id/K_0 $ of the camera $Q_i = 0,1$ in the far field. At this point, we will make the simplifying assumption that the size of the pixels used to resolve transverse momenta is small enough so that no more than one photon ends up in each pixel in a single run of the experiment, implying $n(k_i,Q_i)\in \{0,1\}$. We can then concisely label any $L$-photon outcome as a composite specification of the number of photons detected in each camera $(X = \sum_{i}Q_i,\, L-X)$, and a consolidated list of all photonic momenta $\{k_m\}~\forall~ m\in\{1,\dots,L\}$. The list of triggered photonic transverse momenta $\{k_m\}~\forall~ m\in\{1,\dots,L\}$ need not be specify which camera the photon was detected in, owing to the fact that all detected photons are completely indistinguishable.

In order to evaluate interference sampling probabilities of $L$-photon detection $P(X;\{k_m\})$, we will first evaluate the $L$-th order correlation function,
\begin{equation}
    G^{(L)}_{Q_1,\dots,Q_L}(k_1,\dots,k_L) = \mathrm{Tr} \left[ \rho_d\otimes\ket{1}_0\bra{1} E_{Q_1}^{-}(k_1)\dots E_{Q_L}^{-}(k_L)E_{Q_1}^+(k_1)\dots E_{Q_L}^+(k_L)     \right].
\end{equation}
Using the forms of the image state in Eq.~(\ref{eq:lupopirandolaform}), and the detection operator in Eq.~(\ref{eq:detection_op}), we have the following sum over incoherent contributions from each term in the Fock state expansion of the two-mode thermal image state:
\begin{align}\label{eq:G2Kformula}
    G^{(L)}_{Q_1,\dots,Q_{L}}(k_1,\dots,k_{L}) &= p_0\,\sum_{j=0}^{L-1}~p_{L-1-j}^+\, p_{j}^-~|\langle \mathrm{vac} | E_{Q_1}^+(k_1)\dots E_{Q_{L}}^+(k_{L})|L-1-j;j\rangle_1 \ket{1}_0|^2 \nonumber\noindent\\
    &= p_0\,\sum_{j=0}^{L-1}~p_{L-1-j}^+ p_{j}^-~\bigg|\,\langle\mathrm{vac}|\sum_{S_1,\dots,S_{L}} E_{S_1,Q_1}^+(k_1)\dots E_{S_{L},Q_{L}}^+(k_{L})|L-1-j;j\rangle_1 \ket{1}_0\bigg|^2\nonumber\noindent\\
    &\equiv T_0 + T_1 + \dots + T_{L-1}.
\end{align}
In the sum over source variables in the $j$-th term, $\langle\mathrm{vac}|\sum_{S_1,\dots,S_{L}} E_{S_1,Q_1}^+(k_1)\dots E_{S_{L},Q_{L}}^+(k_{L})|L-1-j;j\rangle_1 \ket{1}_0$, there are only $L$ non-vanishing terms for $S_i=0, S_{j\neq i}=1; \,\,i,j\in\{1,\dots,L\}$, each term corresponding to the annihilation of the single-photon reference state in input $S_i=0$, and $L-1$ thermal photons in the source input $S_{j\neq i}=1$. Using the commutation relations between the symmetric/antisymmetric mode operators and momentum space operators for the source input,
\begin{align}\label{eq:commplus}
    [a_1(k),a_+^{\dag}]&= \frac{1}{\sqrt{2\pi}}\,\int dx_1 dx_2 ~e^{-ikx_1} \left( \frac{\psi(x_2+s/2)+\psi(x_2-s/2)}{\sqrt{2(1+\delta)}}  \right)~[a_1(x_1),a_1^{\dag}(x_2)] = \frac{2\cos (ks/2) \phi(k)}{\sqrt{2(1+\delta)}},
\end{align}
and similarly
\begin{equation}
    [a_1(k),a^{\dag}_-] = \frac{2i\sin(ks/2)\phi(k)}{\sqrt{2(1-\delta)}},
\end{equation}
the expectation value takes the following form,
\begin{align}
    &\langle\mathrm{vac}|\sum_{S_1,\dots,S_{L}} E_{S_1,Q_1}^+(k_1)\dots E_{S_{L},Q_{L}}^+(k_{L})|L-1-j;j\rangle_1 \ket{1}_0&\nonumber\noindent\\
    &~~~~~~~~~~~~~~~~~~~~~~~~~~~~~~~~~~~~~~~~~~~~= \prod_{m=1}^L~\phi(k_m) \, \sqrt{(L-1-j)!\,j!}  \frac{1}{(1+\delta)^{\frac{L-1-j}{2}}(1-\delta)^{\frac{j}{2}}}~\sum_{i=1}^L \frac{e^{i\phi_i}}{\sqrt{2}} \xi_j(k_1,\dots,k_{i-1},k_{i+1},\dots,k_L), 
\end{align}
where 
\begin{equation}
    \xi_j(k_1,\dots,k_{L-1}) = \sum_{\substack{i_1,\dots,i_{L-1}\in \{1,\dots,L-1\} \\ \mathrm{all\,perms}}}~\cos\left(\frac{k_{i_1}s}{2}\right) \dots \cos\left(\frac{k_{i_{L-1-j}}s}{2}\right) \sin\left(\frac{k_{i_{L-j}}s}{2}\right)\dots \sin\left(\frac{k_{i_{L-1}}s}{2}\right)
\end{equation}
is the trigonometric function that confers the expectation value its oscillatory dependence on the separation $s$, and will be critical in characterising the metrological power of the various orders of correlation functions $G^{(L)}$ in resolving the thermal sources. Finally,
\begin{equation}
    \phi_i = \Phi(1,Q_1) + \dots + \Phi(0,Q_i) + \dots \Phi(1,Q_L)
\end{equation}
are the interferometric phases resulting from the propagation of $L-1$ source photons and the single-photon reference state to the transverse momenta-resolving pixels on the output side of the BS. The $j$-th term then has the following final form
\begin{align}
    T_j = \frac{p_0}{2} \frac{N_s^{L-1}}{(1+N_s(1+\delta))^{L-j-1} (1+N_s(1-\delta))^{j}} \prod_{m=1}^L |\phi(k_m)|^2 ~\bigg| \sum_{i=1}^L e^{i\phi_i} \xi_j(k_1,\dots,k_{i-1},k_{i+1},\dots,k_L )   \bigg|^2
\end{align}
The output probability for $L$-photon detection event $(X;k_1,\dots,k_L)$ is given by the final form
\begin{equation}
    P^{(L)}(X =\sum_{m}Q_m;\{k_m\}) = \frac{1}{2}\prod_{m=1}^{L}|\phi(k_m)|^2~\sum_{j=0}^{L-1} \frac{ \Theta_j p_0N_s^{L-1}}{(1+N_s(1+\delta))^{L-j-1}(1+N_s(1-\delta))^{j}}~\bigg| \sum_{i=1}^L e^{i\phi_i} \xi_j(k_1,\dots,k_{i-1},k_{i+1},\dots,k_L )   \bigg|^2
\end{equation}
where the factor $\Theta_j = \frac{(L-1-j)!j!}{X!(L-X)!}$ compensates for the indistinguishability of the $L$ photons at the output, from either the thermal electromagnetic sources or the single-photon reference, and avoids double-counting the same physical event twice, yielding the final form of Eq.~(\ref{eq:PLform}).  

\section{Coincidence Probabilities for $L=3,4$}\label{app:higherorder_prob}
\subsection{$L=3$}
Multiphoton interference resulting in detection of three photons at the output BS ports may occur in two non-trivially distinct ways -- completely bunched so \emph{all} photons end up in the same camera -- $X=3\,\mathrm{or}\,0$~(B), and unbalanced anti-bunched output in which two photons end up in the same camera while the other photon is detected in the other camera -- $X=1\,\mathrm{or}\,2$~(UA). The corresponding probability for each outcome may be obtained by setting $L=3$ in Eq.~(\ref{eq:PLform}) and collecting terms,
\begin{align}\label{eq:threephotonprob_B}
    P^{(3)}(B;k_1,k_2,k_3) &= P^{(3)}(3;k_1,k_2,k_3) +  P^{(3)}(0;k_1,k_2,k_3)\nonumber\noindent\\
    &=f^{(3)}(B) \prod_{m=1}^3 |\phi(k_m)|^2\sum_{j=0}^2 \Xi^{(3)}_j \bigg| \xi_j(k_2,k_3) + \Lambda_1^{(3)}(B)\,\xi_j(k_1,k_3) + \Lambda_2^{(3)}(B)\, \xi_j(k_1,k_2) \bigg|^2,
\end{align}
and 
\begin{align}\label{eq:threephotonprob_UA}
    P^{(3)}(\mathrm{UA};k_1,k_2,k_3) &= P^{(3)}(2;k_1,k_2,k_3) +  P^{(3)}(1;k_1,k_2,k_3)\nonumber\noindent\\
    &=f^{(3)}(\mathrm{UA}) \prod_{m=1}^3 |\phi(k_m)|^2\sum_{j=0}^2 \Xi^{(3)}_j \bigg| \xi_j(k_2,k_3) + \Lambda_1^{(3)}\,(\mathrm{UA})\xi_j(k_1,k_3) + \Lambda_2^{(3)}\,(\mathrm{UA}) \xi_j(k_1,k_2) \bigg|^2,
\end{align}
where 
\begin{equation}
    \Xi^{(3)}_0 = \frac{2p_0N_s^2}{(1+N_s(1+\delta))^2}; ~~~ \Xi^{(3)}_1 = p_0^2 N_s^2; ~~~ \Xi^{(3)}_{2} = \frac{2p_0N_s^2}{(1+N_s(1-\delta))^2}
\end{equation}
are functions of the source parameters $N_s$ and $\delta$ only, while the factors $f^{(3)}(X)$ (which results from the fact that the collected photons at the output ports are indistinguishable) and $\Lambda_i^{(3)}$ (which is a parity factor that results  from the multiphoton interference at the BS), listed in Table \ref{table:threephotonprob_table}, depend only on the outcome $X$.

\begin{table}[h!]
\centering
\begin{tabular}{|c |c |c |c|} 
 \hline
  &~ $\Lambda_1^{(3)}(X)$~ & ~$\Lambda_2^{(3)}(X)$~ &~ $f^{(3)}(X)$~ \\  
 \hline
 $\mathrm{X=B}$ &+1 & +1 & 1/6 \\ 
 \hline
 $\mathrm{X=UA}$ & +1 & -1 & 1/2 \\
 \hline
\end{tabular}
\caption{List of outcome dependent factors used to construct outcome probabilities for three-photon coincident detection $P^{(3)}(k_1,k_2,k_3)$ defined in Eq.~(\ref{eq:threephotonprob_B}) and (\ref{eq:threephotonprob_UA}).}
\label{table:threephotonprob_table}
\end{table}
In a manner similar to the case of two-photon detection i, we see from the form in Eqs.~(\ref{eq:threephotonprob_B}) and (\ref{eq:threephotonprob_UA}) that the coincidence probabilities for the two possible outcomes for $L=3$ are oscillating functions of the detected transverse momenta, with periods of oscillation that depend on the separation $s$, allowing for a similar estimation of $s$ from sampled three-photon outcomes. From the form of the three-photon output probability in Eqs.~(\ref{eq:threephotonprob_B}) and (\ref{eq:threephotonprob_UA}), and the parities $\Lambda_i^{(3)}$ in Table \ref{table:threephotonprob_table}, we also note that there is no way in which one may construct three-photon events whose probabilities vanish for $s/\sigma_x\rightarrow 0$, a consequence of the extended Hong-Ou-Mandel effect~\cite{alsing2022extending,alsing2025examination} at work here that precludes such ``rare" events when it is impossible to distribute an \emph{equal} number of photons in the two output ports.

\subsection{$L=4$}
Multiphoton interference resulting in detection of four photons at the output ports of the BS may occur in three non-trivially distinct outcomes -- completely bunched so \emph{all} photons end up in the same camera -- $X=4\,\mathrm{or}\,0$~(B), unbalanced anti-bunched so three photons end up in one camera while a lone photon is detected in a different camera -- $X=1\,\mathrm{or}\,3$~(UA), and finally, balanced anti-bunched so two photons are detected in each output BS port -- $X=2$~(A). The corresponding probabilities for each of these three outcomes may be evaluated by setting $L=4$ in Eq.~(\ref{eq:PLform}), yielding
\begin{align}\label{eq:fourphotonprob}
    &P^{(4)}(X;k_1,k_2,k_3,k_4) = \nonumber\noindent\\  
     &~~f^{(4)}(X)\prod_{m=1}^4 |\phi(k_m)|^2 \sum_{j=0}^3 \Xi^{(4)}_j\,\bigg| \xi_j(k_2,k_3,k_4) + \Lambda_1^{(4)}(X)\xi_j(k_1,k_3,k_4)+ \Lambda_2^{(4)}(X) \xi_j(k_1,k_2,k_4) + \Lambda_3^{(4)}(X) \xi_j(k_1,k_2,k_3)   \bigg|^2
\end{align}
where
\begin{align}
    \Xi^{(4)}_0 = \frac{6p_0N_s^3}{(1+N_s(1+\delta))^3};\, \Xi_1^{(4)} = \frac{2p_0^2N_s^3}{(1+N_s(1+\delta))};\Xi_2^{(4)} = \frac{2p_0^2N_s^3}{(1+N_s(1-\delta))};\, \Xi_3^{(4)} = \frac{6p_0N_s^3}{(1+N_s(1-\delta))^3}
\end{align}
are functions of the source parameters $N_s$ and $\delta$ only, while factors $f^{(4)}(X)$~(which results from the indistinguishability of detected photons) and $\Lambda_i^{(4)}(X)$~(which is a sign factor resulting from interference at the BS), listed in Table \ref{table:fourphotonprob}, depend exclusively on the outcome $X$.
\begin{table}[h!]
\centering
\begin{tabular}{|c |c |c |c| c|} 
 \hline
  &~ $\Lambda_1^{(4)}(X)$~ & ~$\Lambda_2^{(4)}(X)$~ &~ $\Lambda_3^{(4)}(X)$~ &~ $f^{(4)}(X)$~ \\  
 \hline
 $\mathrm{X=B}$ &+1 & +1 & +1 & 1/24 \\ 
 \hline
 $\mathrm{X=A}$ & +1 & -1 & -1 & 1/8 \\
 \hline
 $\mathrm{X=UA}$ & -1 & -1 & -1 & 1/6 \\
 \hline
\end{tabular}
\caption{List of outcome dependent factors used to construct outcome probabilities for four-photon coincident detection $P^{(4)}(k_1,k_2,k_3,k_4)$ defined in Eq.~(\ref{eq:fourphotonprob}).}
\label{table:fourphotonprob}
\end{table}

In a manner similar to the case of $L=2$ and $L=3$, we see again from the form in Eq.~(\ref{eq:fourphotonprob}) that the coincidence probabilities for the three possible outcomes for four-photon detection are oscillating functions of the detected transverse momenta, with periods of oscillation that depend on the separation $s$, allowing for a similar estimation of $s$ from sampled four-photon outcomes. We also note the markedly different consequence of the Hong-Ou-Mandel effect for four-photon input, where the outcome probability for $X=B$ can be seen to vanish for $s/\sigma_x\rightarrow 0$, following from the structure of the parities $\Lambda_i^{(4)}$ in Table \ref{table:fourphotonprob}. It is this structure of multiphoton probabilities, resulting from the interference at the BS, that yields superresolution for all even-order photon detection events in our scheme.

\section{Two Photon Interference Sampling~($K=2$)}\label{sec:2PFIs}
For the case of two-photon coincident detection, the FI may be decomposed as the following sum of distinct samplings of variables $X$, $\bar{K}$, and $\Delta k$~(see Appendix \ref{app:decomposition_CFI2P} for details):
\begin{align}
    &F^{(2)} = F^{(2),\mathrm{NR}} + \sum_{X}P^{(2)}(X)\int d\Delta k \left(\frac{\partial \log g(\Delta k;X)}{\partial s}\right)^2 + \sum_{X}P^{(2)}(X)\int d\Delta k \left(\frac{\partial\log f(\bar{K};X)}{\partial s}\right)^2,
\end{align}
which follows from the corresponding decomposition of the two-photon probability density,
\begin{equation}
    P^{(2)}(\bar{K},\Delta k;X) = P(X) f(\bar{K};X) g(\Delta k;X),
\end{equation}
where 
\begin{align}
    &P(X) = \int\int d\bar{K} d\Delta k\,P^{(2)}(\bar{K},\Delta k;X) 
\end{align}
is the probability of detecting bunched or anti-bunched BS output \emph{without} resolving the transverse momenta of the photons, and the resolved conditional probabilities are 
\begin{align}\label{eq:conditional_Kbar}
    f(\bar{K};X) = |\phi(\bar{K})|^2~ \frac{1+N_s-\alpha(X)N_s\delta \cos (\bar{K}s)}{1+N_s-\alpha(X)N_s\delta \kappa_1(s) },
\end{align}
and
\begin{align}\label{eq:conditonal_deltak}
    g(\Delta k;X) = C(\Delta k) ~\frac{1+\alpha(X)\cos(\Delta ks/2)}{1+\alpha(X)\kappa_2(s)},
\end{align}
where $\kappa_1(s) = \int |\phi(\bar{K})|^2 \cos(\bar{K}s) d\bar{K}$ and $\kappa_2(s) = \int C(\Delta k) \cos(\frac{\Delta ks}{2}) d\Delta k$ are factors that depend on the PSF $\psi(x)$; for Gaussian PSFs are simply $\kappa_1 = \kappa_2 = e^{-s^2\sigma_k^2/4}$. 

The structure of the conditional probabilities in Eqs.~(\ref{eq:conditional_Kbar}) and (\ref{eq:conditonal_deltak}) reveals the relative importance of the sampling in $\bar{K}$ and $\Delta k$ variables -- following the discussion in Section \ref{sec:twophoton_probs}, we see that the $\delta$-dependence of the amplitude of damped oscillations in $\bar{K}$ probabilities means that $f(\bar{K};X)$ only weakly depends on the separation~($\frac{\partial \log f(\bar{K};X)}{\partial s} \approx 0$), illustrated in Figure \ref{fig:gf_figure} for exemplary values of $s$. The precision of the imaging scheme is consequently controlled by samplings in variables $\Delta k$ and $X$ only,
\begin{equation}
    F^{(2)} \approx F^{(2)}_{\Delta k;X} 
\end{equation}
where $F^{(2)}_{\Delta k;X}$ is the FI that sets sensitivity bounds on samplings of $(\Delta k,X)$ only, defined in Appendix \ref{app:decomposition_CFI2P}. This is borne out in Figure \ref{fig:twophoton_FIs} (a), where a complete sampling of both $\Delta k$ and $\bar{K}$,~(along with which-detector measurement $X$) is seen to be numerically equivalent to a sampling of $\Delta k$ only, for all separations $s$. 

Finally, we note that the relative importance of the momentum resolution~(specifically $\Delta k$ resolution) changes with source separation. For sub-Rayleigh separations $s/\sigma_x\ll 1$, $X$-only sampling using bucket detection suffices, as illustrated by the overlap of $X$-sampling CFI $F^{(2)}_{X}$ with the total CFI $F_t^{(2)}$ for small $s/\sigma_x$ in Figure \ref{fig:twophoton_FIs}(a). On the other hand, momentum resolution becomes increasingly important for separations $s/\sigma_x
>1$, leading to a finite precision for the scheme for very large separations, where $X$-sampling is seen to make little contribution. This highlights the functional utility of the transverse momenta resolution, conferring precision to our scheme even when there is little overlap between the PSFs for the two sources. Jointly, the two samplings confer utility to the imaging setup over a range of separations, in contrast with pix-SPADE~\cite{tsang2016quantum,tsang2019resolving} and pix-SLIVER~~\cite{tsang2019resolving,schodt2023tolerance},where estimator variances diverge for large $s/\sigma_x$~(see Appendix \ref{app:2p_largesep} for more detailed discussion).  

\begin{figure}
    \centering
    \includegraphics[width=0.5\textwidth]{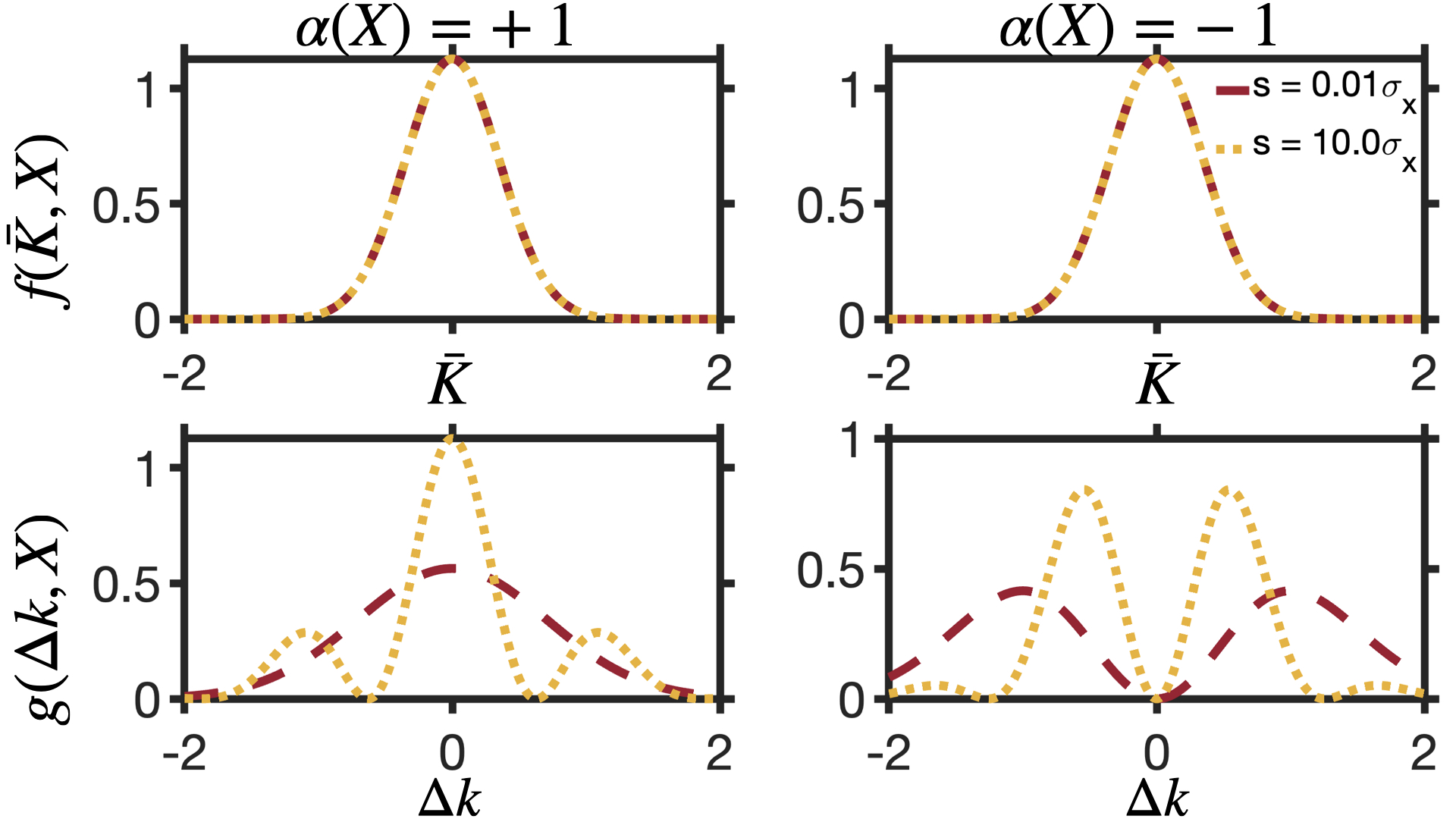}
    \caption{Sensitivity of resolved conditional probabilities $f(\bar{K};X)$~(corresponding to $\bar{K}$ sampling) and $g(\Delta k;X)$ (for $\Delta k$ sampling) with respect to the normalised separation $s/\sigma_x$ for bunched~($\alpha(X)=+1$) and antibunched~($\alpha(X) = -1$) output. Red dashed lines correspond to $s/\sigma_x = 0.01$, and yellow dots correspond to $s/\sigma_x = 10.0$.}
    \label{fig:gf_figure}
\end{figure}

\begin{figure*}
    \centering
    \includegraphics[width=0.44\textwidth]
    {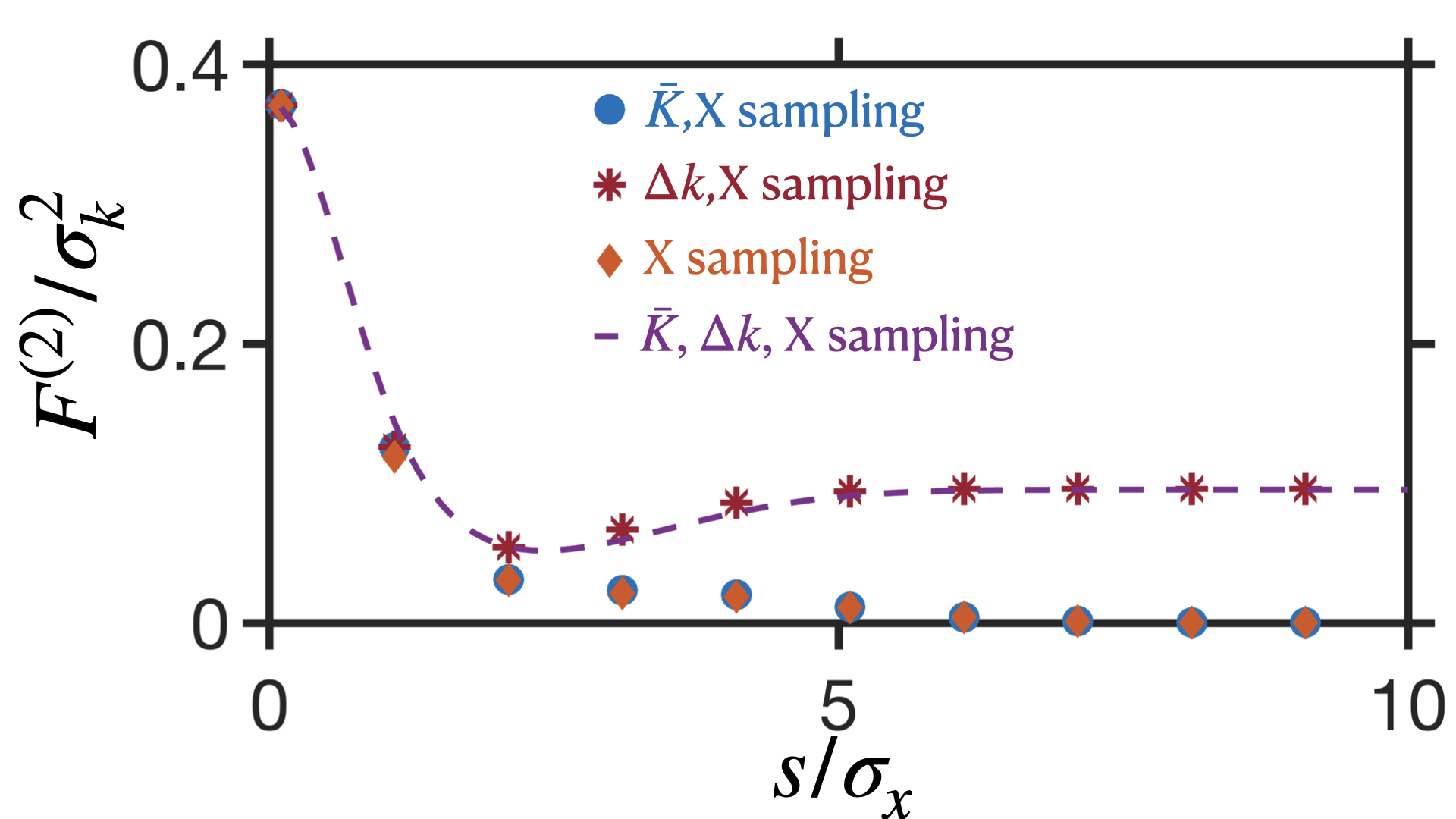}
    \includegraphics[width=0.44\textwidth]{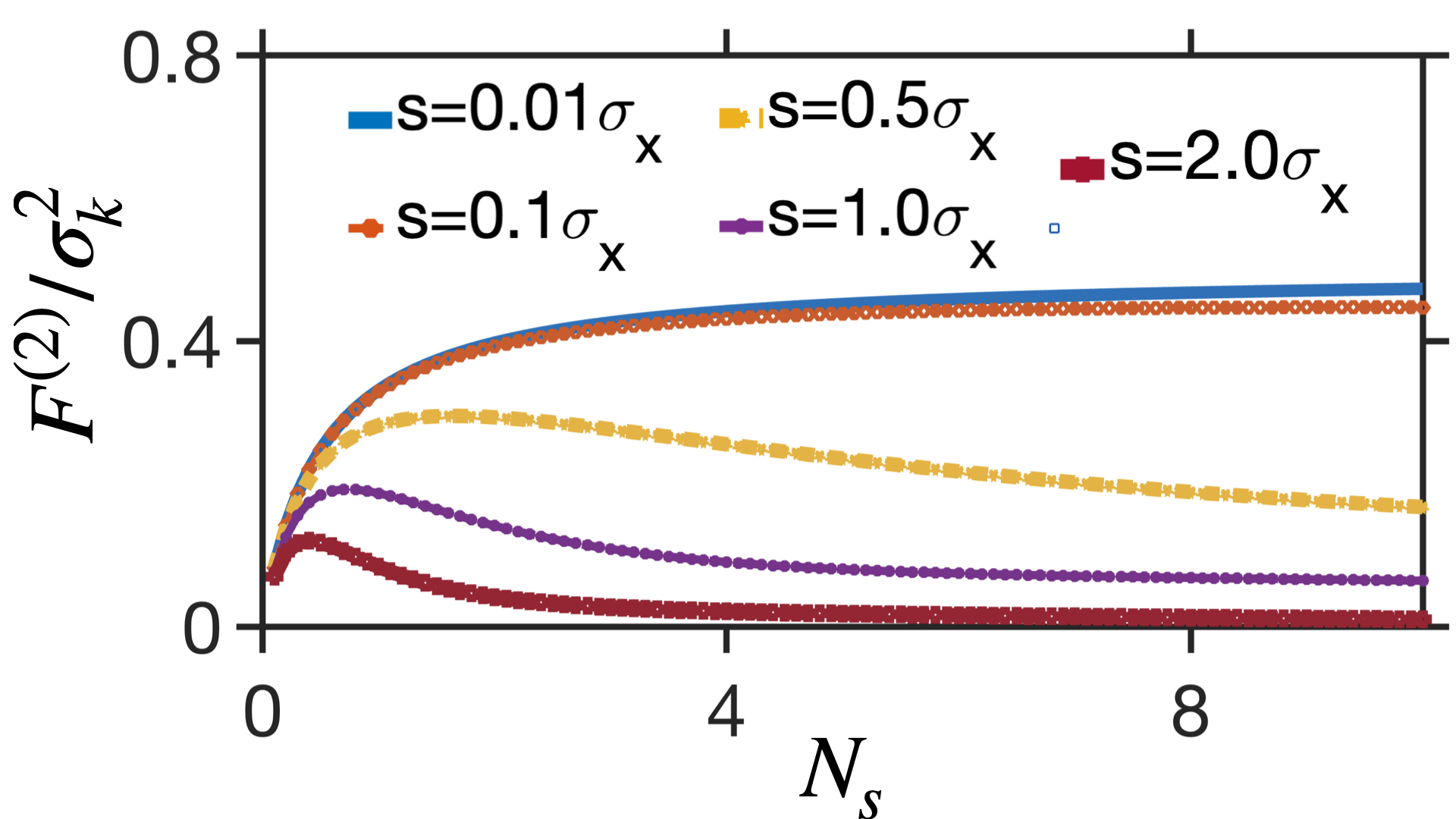}
\caption{(a) Two-photon FI component $F^{(2)}/\sigma_k^2$ for two-photon sampling measurements as a function  of normalised separations $s/\sigma_x$, for the hierarchy of samplings in variables $(\Delta k, \bar{K},X)$.~($N_s=1.5$) (b) FI component $F^{(2)}/\sigma_k^2$ as a function of source strength $N_s$ for a series of values of normalised separation $s/\sigma_x$. All figures correspond to a Gaussian PSF $\psi(x)$ with $\sigma_x = 1.0$. }
    \label{fig:twophoton_FIs}
\end{figure*}

\emph{$N_s$ scaling and optimal brightness:} Next, we note from Figure \ref{fig:twophoton_FIs} (b) the $N_s$-scaling of the total two-photon FI $F_t^{(2)}$ as source separation $s$ changes. For weak to moderate intensities corresponding to $N_s<1$, the FI~(and hence corresponding estimator precisions) increases linearly with respect to $N_s$ for all $s$, with a slope directly proportional to transverse momentum spread $\sigma_k^2$. This corresponds to increasing (collective) single-photon emission from the source pair as the brightness $N_s$ increases. This linear scaling also matches the standard quantum limit~(SQL) scaling of the quantum FI $\mathcal{Q}(s/\sigma_x\rightarrow 0) = 2N_s\sigma_k^2$~\cite{tsang2019resolving,lupo2016ultimate}. 

Beyond $N_s>1$, the $N_s$-scaling strongly depends on the magnitude of the source separation $s$. For sub-Rayleigh separations, the FI is given by the analytical expression~(see Appendix \ref{app:subRayleigh_app} for details),
\begin{equation}\label{eq:superresolution_twophoton}
    F^{(2)}(s/\sigma_x\rightarrow0) = \frac{N_s}{1+2N_s}\sigma_k^2.
\end{equation}
This behaviour with respect to $N_s$ in Eq.~(\ref{eq:superresolution_twophoton}) can be understood as following: in the limit of small separations $s$, the leading order term in $F_t^{(2)}(s)$ arises solely from the anti-bunching probabilities induced by the anti-symmetric two-mode thermal state term $\ket{0;1}$, which in turn is proportional to the probability of the source emitting a photon in the antisymmetric mode~($p_0p_1^{-}$), as well as the squared magnitude of the commutator $\bigg|[a_1(k_1),a_+^{\dag}] - [a_1(k_2),a_+^{\dag}]\bigg|^2$ which represents the interference term between the overlap of the antisymmetric photon with detected photons with fixed transverse momenta $k_1$ and $k_2$~(the negative sign arises from the phases induced by antibunching). The consequent sub-SQL scaling of the FI $F_t^{(2)}(s)$ tracks the sub-SQL scaling of the weight of the anti-symmetric component of the source states, tending to the finite value of $\sigma_k^2/2$ for $N_s\rightarrow\infty$.

In contrast, for large separations $s/\sigma_x \gg 1$, the  FI scales as the probability of the collective emission of one photon from the source~(see Appendix \ref{app:2p_largesep}),
\begin{equation}
    F^{(2)}_{\mathrm{asymptotic}} = \frac{N_s}{(1+N_s)^3}\sigma_k^2,
\end{equation}
so that the asymptotic separation FI peaks  at $N_s=0.5$ when collective one-photon emission is most likely. 

This physically motivated reasoning can be extended to multiphoton detection beyond $L>2$, so that $ \mathrm{argmax}_{N_s}~  F^{(L)}_{\mathrm{asymptotic}} = \frac{L-1}{2}$, corresponding to maximum likelihood of the source emitting $L-1$ photons that the experiment detects, in addition to the single-photon reference, in the output cameras. This reasoning also offers a helpful guide to source engineering in our scheme for separation estimation of artificial thermal sources whose brightness is controllable -- for large enough separation, the source brightness can be tuned to the number of photons that \emph{can} be detected in coincidence. On the other hand, if the source brightness is fixed, coincidence detection of $L-1$ photon events will yield most information.

\section{Decomposition of Sampling CFIs for Two-Photon Detection}\label{app:decomposition_CFI2P}
Total probability of detecting two photons with transverse momenta difference $\Delta k = k_1-k_2$ and mean $\bar{K} = \frac{k_1+k_2}{2}$ in bunching~($X=+1$) or antibunching event~($X=-1$) is obtained by setting $L=2$ in Eq.~(\ref{eq:PLform}):
\begin{equation}
    P^{(2)}(\Delta k,\bar{K};X) = N_sp_0^2~|\phi(\bar{K})|^2C(\Delta k)~ (1+N_S-\alpha(X)N_s \delta \cos (\bar{K}s)) \bigg[1+\alpha(X)\cos\left(\frac{\Delta ks}{2}\right)\bigg]
\end{equation}
Marginal probabilities of reduced sampling in a smaller number of variables can then be obtained as
\begin{align}
    P^{(2)}(\Delta k;X) = \int d\bar{K}~ P^{(2)}(\Delta k,\bar{K};X) = N_s p_0^2~C(\Delta k)~(1+N_s-\alpha(X)N_s\delta \kappa_1(s)) \bigg[1+\alpha(X)\cos\left(\frac{\Delta ks}{2}\right)\bigg],
\end{align}
\begin{align}
    P^{(2)}(\bar{K};X) = \int d\Delta k~ P^{(2)}(\Delta k,\bar{K};X) = N_s p_0^2~|\phi(\bar{K})|^2~(1+N_s-\alpha(X)N_s\delta \cos(\bar{K}s)) \bigg[1+\alpha(X)\kappa_2(s)\bigg],
\end{align}
and
\begin{align}
    P^{(2)}(X) = \int d\bar{K}\int d\Delta k ~ P^{(2)}(\Delta k,\bar{K};X) = N_sp_0^2 (1+N_s-\alpha(X)N_s\delta \kappa_1(s))\,\bigg[ 1+\alpha(X)\kappa_2(s)  \bigg]
\end{align}
where $\kappa_1(s) = \int |\phi(\bar{K})|^2 \cos(\bar{K}s) d\bar{K}$ and $\kappa_2(s) = \int C(\Delta k) \cos(\frac{\Delta ks}{2}) d\Delta k$ are factors that depend on the PSF $\psi(x)$; for Gaussian PSFs are simply $\kappa_1 = \kappa_2 = \mathrm{exp}\left(-\frac{s^2\sigma_k^2}{4}\right)$. 

Each of the probability distributions correspond to a sampling that is sensitive to only the parameters in the arguments -- respectively, the difference in the detected momenta  and which-detector measurement~($P^{(2)}(\Delta k;X)$), the average detected momenta and which-detector measurement~($P^{(2)}(\bar{K};X)$), and finally only a which-detector measurement that does not resolve in the transverse momenta at all~($P^{(2)}(X)$). In terms of the which-detector only probability, the other marginal probability densities can be re-expressed as product decompostion using the functions,
\begin{equation}
    f(\bar{K};X) = |\phi(\bar{K})|^2 ~ \frac{1+N_s - \alpha(X)N_s\delta \cos(\bar{Ks})}{1+N_s-\alpha(X)N_s\delta e^{-s^2\sigma_k^2/4}}; ~~\int d\bar{K}~f(\bar{K};X) = 1 ~\forall ~ X=A,B.
\end{equation}
and
\begin{equation}
    g(\Delta k;x) = C(\Delta k) \frac{1+\alpha(X)\cos\left(\frac{\Delta ks}{2}\right)}{1+\alpha(X) e^{-s^2\sigma_k^2/4}}; ~~\int d\Delta k g(\Delta k;X) = 1 ~\forall~X=A,B.
\end{equation}
which gives for the total sampling probability,
\begin{equation}
    P^{(2)}(\Delta k,\bar{K};X) = P^{(2)}(X) f(\bar{K};X) g(\Delta k;X);
\end{equation}
and reduced samplings
\begin{equation}
    P^{(2)}(\Delta k;X) = P^{(2)}(X) g(\Delta k;X),~~~P^{(2)}(\bar{K};X) = P^{(2)}(X)f(\bar{K};x)
\end{equation}
The score function can be then expressed as the following for the above probability distributions:
\begin{align}
    &\frac{1}{P^{(2)}(\Delta k,\bar{K};X)}~\left( \frac{\partial P^{(2)}(\Delta k,\bar{K};X)}{\partial s} \right)^2 \nonumber\noindent\\
    &= f(\Delta K;X) g(\bar{K};X)~\frac{1}{P^{(2)}(X)}\left( \frac{\partial P^{(2)}(X)}{\partial s}\right)^2 + P^{(2)}(X)f(\bar{K};X)~\frac{1}{g(\Delta k;X)}\left( \frac{\partial g(\Delta k;X)}{\partial s} \right)^2 \nonumber\noindent\\
    &+ P^{(2)}(X)g(\Delta k;X)~\frac{1}{f(\bar{K};X)}\left( \frac{\partial f(\bar{K};X)}{\partial s} \right)^2 + 2g(\Delta k;X) \frac{\partial P^{(2)}(X)}{\partial s} \frac{\partial f(\bar{K};X)}{\partial s} + 2P^{(2)}(X)\frac{\partial f(\bar{K};X)}{\partial s}\frac{\partial g(\Delta k;X)}{\partial s} \nonumber\noindent\\
    &+2f(\bar{K};X) \frac{\partial P^{(2)}(X)}{\partial s} \frac{\partial g(\Delta k;X)}{\partial s}.
\end{align}
Using the fact that the variation of integrals of probability distributions vanish $\int d\bar{K}~\frac{\partial}{\partial s} f(\bar{K};X)  = \int d\Delta k \frac{\partial}{\partial s} g(\Delta k;X)  = 0$, we have 
\begin{align}
    F^{(2)} &= \int d\bar{K}\int\Delta k \sum_X \frac{1}{P^{(2)}(\Delta k,\bar{K};X)}~\left( \frac{\partial P^{(2)}(\Delta k,\bar{K};X)}{\partial s} \right)^2  \nonumber\noindent\\
    &= F_{\mathrm{X}} + \sum_X P^{(2)}(X) \int d\Delta k ~\frac{1}{g(\Delta k;X)}\left( \frac{\partial g(\Delta k;X)}{\partial s} \right)^2 + \sum_X P^{(2)}(X) ~\frac{1}{f(\bar{K};X)}\left( \frac{\partial f(\bar{K};X)}{\partial s} \right)^2 \nonumber\noindent\\
    &=F_{\mathrm{X}} + \Delta F_{\Delta k} + \Delta F_{\bar{K}}.
\end{align}
Similarly, we can decompose the FIs corresponding to reduced samplings of the variables as 
\begin{align}
    F^{(2)}_{\Delta k;X} &= F_{\mathrm{X}} + \Delta F_{\Delta k}; \nonumber\noindent\\
    F^{(2)}_{\bar{K};X} &= F_{\mathrm{X}} + \Delta F_{\bar{K}};\nonumber\noindent\\
    F^{(2)}_{\mathrm{NR}} &= F_{\mathrm{X}}
\end{align}
establishing the hierarchy
\begin{equation}
    F^{(2)} \geq F^{(2)}_{\Delta k;X(\bar{K};X)} \geq F_{\mathrm{NR}}.
\end{equation}

\section{Two-Photon FI $F^{(2)}(s)$ for $s/\sigma_x\rightarrow \infty$}\label{app:2p_largesep}
In the limit of $s/\sigma_x\rightarrow \infty$, the two-photon probability of detecting photons with transverse momenta difference $\Delta k = k_1-k_2$ and mean $\bar{K} = \frac{k_1+k_2}{2}$ is given as 
\begin{equation}
    P^{(2)}_{\mathrm{asymptotic}}(\Delta k,\bar{K};X) = \frac{N_s}{(1+N_s)^3} ~C(\Delta k)|\phi(\bar{K})|^2~ ~\bigg[1+\alpha(X)\cos\left( \frac{\Delta ks}{2} \right)\bigg]
\end{equation}
where we have imposed the limit
\begin{equation}
    \lim_{s/\sigma_x\rightarrow 0}  N_s\delta\cos(\bar{K}s) =0 
\end{equation}
in Eq.~(\ref{eq:G2_probability}). We note that $P^{(2)}_{\mathrm{asymptotic}}(\Delta k,\bar{K};X)$ is independent of $\bar{K}$, revealing the separation-damped amplitude of $\bar{K}$-beating that dies out as the source separation increases. The score function is then given as 
\begin{equation}
    \frac{1}{P^{(2)}_{\mathrm{asymptotic}}(\Delta k,\bar{K};X)} \left( \frac{\partial P^{(2)}_{\mathrm{asymptotic}}(\Delta k,\bar{K};X)}{\partial s } \right)^2 = \frac{N_s}{(1+N_s)^3}~C(\Delta k) |\phi(\bar{K})|^2 ~\frac{(\Delta k^2/4)\,\sin^2 \frac{\Delta k s}{2}}{1+\alpha(X)\cos \frac{\Delta k s}{2} }
\end{equation}
The asymptotic separation CFI is then Gaussian PSFs:
\begin{align}\label{eq:C2asymp}
    F^{(2)}_{\mathrm{asymptotic}} &= \sum_{X=+1,-1} \int d\Delta k \int d\bar{K} \frac{1}{P^{(2)}_{\mathrm{asymptotic}}(\Delta k,\bar{K};X)} \left( \frac{\partial P^{(2)}_{\mathrm{asymptotic}}(\Delta k,\bar{K};X)}{\partial s } \right)^2 \nonumber\noindent\\
    &= \frac{1}{2} \frac{N_s}{(1+N_s)^3} \left( \sigma_k^2 + \sigma_k^2  \right) = \frac{N_s \sigma_k^2}{(1+N_s)^3},
\end{align}
where we note that in the limit of very large separations, the FI contribution of bunched~($X=+1$) and anti-bunched events are equal, and half the asymptotic FI $F^{(2)}_{\mathrm{asymptotic}}$, a feature that was observed for interferometric inner-variable sampling schemes in \cite{triggiani2024estimation}.

We also remark that the scaling of the asymptotic FIs with respect to source brightness $N_s$ in Eq.~(\ref{eq:C2asymp}) is precisely the scaling of collective one-photon emission events from the thermal sources, increasing linearly for weak sources, peaking at $N_s=0.5$, and subsequently decaying to $0$. This implies that the two-photon detection itself offers an $N_s$-independent precision for large separations. A more intuitive understanding is gained by recalling that for the PSF overlap $\delta\rightarrow 0$, the two-photon detection images the (renormalised) state
\begin{equation}
    \rho_{\mathrm{2P,conditonal}} = \frac{1}{2}~\bigg (\ket{1;0}\bra{1;0} + \ket{0;1}\bra{0;1}\bigg)
\end{equation}
where $\ket{1;0}$ and $\ket{0;1}$ are orthogonal (image) spatial modes corresponding to the two sources, the orthogonality a consequence of the vanishing PSF overlap. Weighted by the probability of the two sources collectively emitting one photon only~(the other photon in the detection contributed by the single-photon reference) $P_1 = \lim_{\delta\rightarrow 0}\,p_0(p_1^+ + p_1^-) = \frac{2N_s}{(1+N_s)^3}$, the FI of separation is then retrieved 
\begin{equation}
    F^{(2)}_{\mathrm{asymptotic}} = p_0~[p_1^+(\delta \rightarrow 0) + p_1^-(\delta \rightarrow 0)] \times F\bigg[ \rho_{\mathrm{2P,conditional}}  \bigg] = \frac{N_s\sigma_k^2}{(1+N_s)^3},
\end{equation}
where $F\bigg[ \rho_{\mathrm{2P,conditional}}\bigg] = \sigma_k^2/2$~\cite{muratore2025superresolution}.

\section{Sub-Rayleigh form of $P^{(L)}(\{k_i,C_i\})$}\label{app:subRayleigh_app}
In order to establish the superresolution inherent in the multiphoton interference that the imaging scheme leverages in order to resolve thermal sources of arbitrary strength, let us first examine the dependence against source separation $s$ of the superposition term 
\begin{equation}
    \mathcal{I}_j = \bigg| \sum_{i=1}^L e^{i\phi_i} \xi_j(k_1,\dots,k_{i-1},k_{i+1},\dots,k_L \bigg|^2
\end{equation}
for $j=0$, which from the form of the corresponding trigonometric function $\xi_0(\{k_m\}) = \prod_m \cos\left( \frac{k_ms}{2}\right)$ can be seen to be $\mathcal{I}_0 = O(1)$ in $s$ for all outcomes $X$, \emph{except} $X=L/2$ for even-order $L$, for which $\phi_{i}=+1, ~i=1,\dots,L/2$ and $\phi_i=-1~~i=L/2+1,\dots,L$, and the resulting scaling is $\mathcal{I}_0 = O(s^4)$. Recalling the discussion in \ref{section:subrayleigh_seps}, we can see that the corresponding FI contribution stemming from $\mathcal{F}[\mathcal{I}_0] = O(s^2)$, vanishing for small separations and offering no superresolution in the scheme. 

Next, we examine the behaviour of $\mathcal{I}_j$ for $j>1$, for which $\mathcal{\xi}_j \approx \sum_{\mathrm{all\,\, perms}} \left(\frac{k_{i_1}\dots k_{i_j}}{2^j}\right)s^j $ for all orders $L$ and outcomes $X$, implying $\mathcal{I}_j = O(s^j)$. Crucially then, for the choice of even number of detected photons $L=2P$ and balanced anti-bunched output events $X=L/2=P$, the leading order contribution to the output probability stems from the term corresponding to $j=1$, and not $j=0$ as it would for all multiphoton interference outcomes that are not balanced antibunched,
\begin{equation}
    P^{(L=2P)}(X=P;k_1,\dots,k_{2K})\approx  \frac{(2P-2)!)}{P!P!}\frac{N_s^{2P-1}}{(1+2N_s)^{2P-1}} \prod _{m=1}^{2P} |\phi(k_m)|^2~ (k_1+\dots k_{P}-k_{P+1}-\dots -k_{2P})^2~\frac{s^2}{4} + O(s^4),
\end{equation}
from which we can quite easily derive the form of the imaging scheme FI in the sub-Rayleigh limit,
\begin{equation}
    F^{(L=2P)}(s/\sigma_x\rightarrow 0) = \frac{1}{2}\frac{\binom{2P}{P}}{2P-1} ~\left( \frac{N_s}{1+2N_s}\right)^{2P-1}~\sigma_k^2,
\end{equation}
which recovers the form in Eq.~(\ref{eq:fisher_subrayleigh}).

\end{widetext}

\bibliography{bibliography}
\end{document}